\documentclass[sigconf]{acmart}
\copyrightyear{2022}
\acmYear{2022}
\setcopyright{acmcopyright}\acmConference[CCS '22]{Proceedings of the 2022 ACM SIGSAC Conference on Computer and Communications Security}{November 7--11, 2022}{Los Angeles, CA, USA}
\acmBooktitle{Proceedings of the 2022 ACM SIGSAC Conference on Computer and Communications Security (CCS '22), November 7--11, 2022, Los Angeles, CA, USA}
\acmPrice{15.00}
\acmDOI{10.1145/3548606.3560561}
\acmISBN{978-1-4503-9450-5/22/11}

\usepackage{hyperref}
\usepackage[hyphenbreaks]{breakurl}
\usepackage{balance}
\usepackage{url}
\usepackage{color}
\usepackage{caption}
\usepackage{diagbox}
\usepackage{subfigure}
\usepackage{multirow}
\usepackage{booktabs}
\usepackage{epsfig,endnotes}
\usepackage{enumitem}
\newcommand{\RNum}[1]{\uppercase\expandafter{\romannumeral #1\relax}}

\newcommand{\model}{{\mathcal{F}_{\theta}}}

\newcommand{\para}[1]{{\vspace{2pt} \noindent \textbf{#1}
    \hspace{6pt}}}

\newcommand{\emily}[1]{{\color{black} #1}}

\definecolor{applegreen}{rgb}{0.55, 0.71, 0.0}

\newcommand{\rev}[1]{{\color{black} #1}}

\newcommand{\htedit}[1]{{\color{black} #1}}

\newcommand{\bencheck}[1]{{\color{black} #1}}

\newcommand{\etal}{{\em et al.\ }}
\newcommand{\eg}{{\em e.g.,\ }}
\newcommand{\ie}{{\em i.e.,\ }}

\newcommand{\secspace}{\vspace{-0.05in}}

\newcommand{\system}{{\em Neo}}

\newcommand{\ad}[1]{{$\mathcal{A}$}}
\newcommand{\service}[1]{{$\mathcal{S}$}}

\newcommand{\ytface}{{\tt YTFace}}
\newcommand{\cifarS}{{\tt CIFAR10}}
\newcommand{\cifar}{{\tt CIFAR10}}
\newcommand{\skin}{{\tt SkinCancer}}

\newcommand{\imagenet}{{\tt ImageNet}}

\newenvironment{packed_itemize}{
\begin{list}{\labelitemi}{\leftmargin=0.5em}
  \setlength{\itemsep}{1pt}
  \setlength{\parskip}{0pt}
  \setlength{\parsep}{0pt}
  \setlength{\headsep}{0pt}
  \setlength{\topskip}{0pt}
  \setlength{\topmargin}{0pt}
  \setlength{\topsep}{0pt}
  \setlength{\partopsep}{0pt}
}{\end{list}}

\begin{document}

\title{Post-breach Recovery: Protection against White-box Adversarial Examples for Leaked DNN Models}

\author{Shawn Shan}
\email{shawnshan@cs.uchicago.edu}
\affiliation{%
  \institution{University of Chicago}
  \city{Chicago}
  \country{USA}
}
\author{Wenxin Ding}
\email{wenxind@uchicago.edu}
\affiliation{%
  \institution{University of Chicago}
    \city{Chicago}
  \country{USA}
}
\author{Emily Wenger}
\email{ewenger@uchicago.edu}
\affiliation{%
  \institution{University of Chicago}
    \city{Chicago}
  \country{USA}
}
\author{Haitao Zheng}
\email{htzheng@cs.uchicago.edu}
\affiliation{%
  \institution{University of Chicago}
    \city{Chicago}
  \country{USA}
}
\author{Ben Y. Zhao}
\email{ravenben@cs.uchicago.edu}
\affiliation{%
  \institution{University of Chicago}
    \city{Chicago}
  \country{USA}
}

\begin{abstract}
  Server breaches are an unfortunate reality on today's Internet. In the
  context of deep neural network (DNN) models, they are particularly harmful, because a
  leaked model gives an attacker ``white-box'' access to generate
  adversarial examples, a threat
  model that has no practical robust defenses. For practitioners
  who have invested years and millions into proprietary DNNs, e.g. medical
  imaging, this seems like an inevitable disaster looming on the horizon.

  In this paper, we consider the problem of {\em post-breach recovery} for
  DNN models. We propose \system, a new system that creates new versions of
  leaked models, alongside an inference time filter that detects and removes
  adversarial examples generated on previously leaked models. 
  The classification surfaces of different model versions are slightly offset
  (by introducing hidden distributions), and \system\/ detects the overfitting of
  attacks to the leaked model used in its generation. We show that across a
  variety of tasks and attack methods, \system\/ is able to filter out
  attacks from leaked models with very high accuracy, and provides strong
  protection (7--10 recoveries) against attackers who repeatedly breach the
  server. \system\/ performs well against a variety of strong adaptive
  attacks, dropping slightly in \# of breaches recoverable, and demonstrates
  potential as a complement to DNN defenses in the wild. 
\end{abstract}



\maketitle

\secspace
\section{Introduction}
\label{sec:intro}

Extensive research on adversarial machine learning has repeatedly demonstrated 
that it is very difficult to build strong defenses against inference time
attacks, {\em i.e.} adversarial examples crafted by attackers with
full (white-box) access to the DNN model. Numerous defenses have been
proposed, only to fall against stronger adaptive attacks.  Some 
attacks~\cite{athalye2018obfuscated,tramer2020adaptive} break
large groups of defenses at one time, while
others~\cite{distillationbroken,magnetbroken,carlini2020partial,he2021feature}
target and break specific
defenses~\cite{distillation,magnet,shan2020gotta}. Two alternative
approaches remain promising, but face significant challenges.  In adversarial
training~\cite{zheng2016improving, mitdefense, zantedeschi2017efficient}, 
active efforts are underway to overcome challenges in high computation
costs~\cite{shafahi2019adversarial,wong2020fast}, limited
efficacy~\cite{rebuffi2021fixing,zhang2019theoretically,gowal2020uncovering,gowal2021improving},
and negative impact on benign classification. Similarly,  certified
defenses offer provable robustness against $\epsilon$-ball
bounded perturbations, but are limited to small $\epsilon$ and do
not scale to larger DNN architectures~\cite{pmlr-v97-cohen19c}.


These ongoing struggles for defenses against white-box attacks have 
significant implications for ML practitioners. Whether DNN models are
hosted for internal services~\cite{xu2021deep,facebookcontent} or as cloud
services~\cite{ribeiro2015mlaas,mlaas}, attackers can get white-box
access by breaching the host infrastructure. Despite billions of dollars spent
on security software, attackers still breach high value servers,
leveraging a wide range of methods from unpatched software vulnerabilities to
hardware side channels and spear-phishing attacks against employees. Given
sufficient incentives, {\em i.e.} a high-value, proprietary DNN model, it is often a
question of when, not if, attackers will breach a server and compromise its
data. Once that happens and a DNN model is leaked,
its classification results can no longer be trusted, since an attacker can
generate successful adversarial inputs using a wide range of white-box attacks. 


There are no easy solutions to this dilemma.  Once a model is leaked, some services,
{\em e.g.} facial recognition, can recover by acquiring new training data
(at additional cost) and training a new model from scratch. 
Unfortunately, even this
may not be enough, as prior work shows that for the same task,
models trained on different datasets or architectures often exhibit 
transferability~\cite{qin2021adversarial,wu2018understanding}, where
adversarial examples computed using one model may 
succeed on another model. More importantly,  for many safety-critical domains
such as medical imaging, building a new training dataset may simply be
infeasible due to prohibitive costs in time and capital. Typically, data
samples in medical imaging must match a specific
pathology, and undergo de-identification under privacy regulations (e.g. HIPAA
in the USA), followed by careful curation and annotation by certified
physicians and specialists. All this adds up to significant time and financial costs.
For example, the HAM10000 dataset includes 10,015 curated images of skin
lesions, and took 20 years to collect from two medical sites in Austria and
Australia~\cite{tschandl2018ham10000}. The Cancer Genome Atlas (TCGA) is a
17 year old effort to gather genomic and image cancer data, at a
current cost of \$500M
USD\footnote{\url{https://www.cancer.gov/about-nci/organization/ccg/research/structural-genomics/tcga/history/timeline}}. 

In this paper, we consider the question:
{\em as practitioners continue to invest significant amounts of time and
  capital into building large complex DNN models (i.e. data acquisition/curation and
  model training), what can they do to avoid losing their investment following an
  event that leaks their model to attackers (e.g. a server breach)?} We
refer to this as the {\bf{\em post-breach recovery}} problem for DNN
services. 



\para{A Metric for Breach-recovery.} Ideally, a recovery system can generate
a new version of a leaked model that restores much of its
functionality, while remaining robust to attacks derived from the leaked
version. But a powerful and persistent attacker can breach a model's host
infrastructure multiple times, each time gaining additional information to
craft stronger adversarial examples. Thus, we propose {\bf { number of
    breaches recoverable (NBR)}} as a success metric for post-breach recovery
systems. NBR captures the number of times a model owner can restore a model's
functionality following a breach of the model hosting server, before
they are no longer robust to attacks generated on leaked versions of the
model.
For example, an NBR of 0 means the model is highly vulnerable
after a single breach (no recovery), while an NBR of 5 means the model can be
breached 5 times before it becomes vulnerable.

\para{Potential Solution: Adversarial-disjoint Ensembles.} While we know of
no prior attempts to address the post-breach recovery problem, the existing
approach that most closely resembles a solution is ``adversarial-disjoint''
ensembles~\cite{yang2021trs,abdelnabi2021s,kariyappa2019improving,yang2020dverge}, a set of mutually non-transferable models where adversarial
examples optimized on one model does not transfer well to others. Despite
recent attempts, progress has been limited, largely due to the fact that removing
transferability between same-task models is a very
challenging problem~\cite{yang2021trs}. 
Later in \S\ref{sec:compare}, we explore this empirically and show
that SOTA ensemble methods~\cite{yang2021trs,abdelnabi2021s,kariyappa2019improving,yang2020dverge}, when adapted for breach recovery, produce solutions with NBR $<$ 1.


\para{Breach Recovery via Identifiable Model Versions.}  This paper describes \system,
a new approach to help restore a DNN's functionality following a model
breach. At a high level, \system\/ works by producing multiple version of a
trained model, where their classification surfaces are shifted subtly, such
that adversarial examples produced by one version are distinguishable from
those computed on another. If a model version $F_i$ is leaked
following a server breach, $F_i$ is retired, and replaced with a different
version $F_j$, along with a filter representing $F_i$. Incoming queries are
tested to determine if they overfit on $F_i$, and if so, they are filtered and marked as
potential attack inputs. Over time, any model that is leaked following
another server breach is also retired and replaced with another
version. All incoming queries are tested against filters of all previous
leaked models to detect adversarial examples. By leveraging the natural
overfitting of an adversarial example to leaked model version(s), \system\/ can
often tolerate up to 10 server breaches (NBR$\sim$10) before an attacker gathers sufficient
data to produce adversarial examples that successfully attack the next model
version \htedit{while bypassing the filters with a reasonable success rate. }




This paper makes five key contributions.
\vspace{-0.03in}
\begin{packed_itemize}
\item We define the post-breach model recovery problem, and introduce NBR
  (\# of breaches recoverable) as a success metric.
\item We introduce \system, a recovery system that generates model versions
  whose classification surfaces contain small, controlled
  differences. \htedit{This is done by pairing hidden data distributions produced using
  GANs with the original training data.}  Thus \system\/ can detect
adversarial examples generated from one or more 
  leaked model versions at inference time with high accuracy.
\item \htedit{We use formal analysis to validate the design of  \system's attack filter, and
  prove a lower bound on the difference in loss between adversarial examples
  generated from a leaked model and their loss on another version.
  Thus our attack filter can distinguish between adversarial and benign
  inputs by comparing loss across versions.}
\item We evaluate \system\/ on tasks ranging from facial recognition,
  object recognition to cancer classification, and show it is able to
  recover from 7 to 10 model breaches while maintaining robustness against
  adversarial examples generated on leaked models.
\item We evaluate \system\/ against a comprehensive set of
  adaptive attacks (7 total attacks using 2 general strategies). Across four
  tasks, adaptive attacks typically produce small drops (<1) in NBR, and
  \system\/ maintains its ability to recover from multiple model breaches.
\end{packed_itemize}

In practice, we expect post-breach recovery systems to operate in complement
with traditional white-box or black-box DNN defenses. They address the
uncommon yet critical event of a model leak, and can be deployed following
evidence of an infrastructure breach, such as warnings by intrusion detection
systems, or evidence of downstream attacks on the model or other server
components via logs or forensic analysis.

\secspace
\section{Background and Related Work}
\label{sec:back}

In this section, we present background and related work on model leakages, adversarial 
example attacks and defenses. 

\vspace{-0.05in}
\subsection{Model Leakage}
\label{sec:back_leakage}


\htedit{Today, DNN models can be hosted on internal servers to answer
internal queries~\cite{xu2021deep,facebookcontent} or external-facing servers as cloud
services (\eg MLaaS~\cite{ribeiro2015mlaas}).  The ``safety'' of these
models depends heavily on the integrity of the hosting server.  A long line of security
research exists to protect remote servers against server breaches.
These include intrusion prevention/detection systems
to detect and block unauthorized server access~\cite{liao2013intrusion,
hoque2012implementation,ids1},  and
human-focused systems that protect employees from spear-phishing
attacks~\cite{jakobsson2005modeling,minkdeepphish} and
strengthen security awareness~\cite{cone2007video}.  Recent
work~\cite{sun2020shadownet,duy2021confidential}  also proposed
methods to securely host ML models leveraging
hardware features such as trusted execution environments (TEE).
}


\htedit{ While these defenses increase the difficulty of breaching remote
  servers~\cite{sec-report}, their protection is still limited.  In fact,
  server breaches are still
  commonplace~\cite{breach-target,breach-eq}, because 
  persistent and resourceful attackers (\eg state-sponsored threat group)
  continue to exploit unpatched vulnerabilities\footnote{Over $200$ critical
    security vulnerabilities are identified in 2020 alone~\cite{sec-report}.}
  and launch more sophisticated attacks to breach even high security
  servers~\cite{mitre}. Beyond software exploits, recent attacks
  exploited supply chains to inject backdoors into source
  code~\cite{jibilian2021us}, while new exploits such as GPU/memory side
  channels offer new ways to steal models~\cite{hu2021systematic,
    hua2018reverse,rakin2021deepsteal}.}

\vspace{-0.05in}
\subsection{Adversarial Example Attacks on DNNs}
\label{subsec:attack}

\htedit{Adversarial examples are an inference time attack, where an adversary
  crafts an imperceptible perturbation ($\delta$) for an input $x$, such that
  the target model $\model$  misclassifies $x+\delta$ to a target label
  $y_t=\model(x+\delta) \ne \model(x)$. }


\htedit{A leaked model following a server breach provides an attacker with
  the strongest possible attack model: {\em white-box}
  access to the model parameters, and the ability to optimize $\delta$ to
  maximize attack success. Below we summarize three
  SOTA white-box adversarial attack methods frequently used to evaluate defenses.}


\begin{packed_itemize}\vspace{-0.02in}
\item \textbf{PGD}~\cite{kurakin2016adversarial} \htedit{crafts adversarial
  perturbation using an iterative
  search guided by signed gradient descent.} Let $x$ be the original input,  $y_t$ the target
  label, and $\delta_{n}$ the adversarial perturbation computed for $x$ at the $n^{th}$ optimization step. Then, $\delta_n = \eta \cdot \text{sign}(\nabla_x \ell(\model(x + \delta_{n-1}), y_t))$
where $\eta$ is the optimization step size and $\delta_n$ is clipped to have 
$L_{inf}$ norm smaller than a designated attack budget. 

\item \textbf{CW}~\cite{cwattack} uses gradient optimization to
search for an adversarial perturbation by minimizing both $L_p$ norm of the perturbation and  attack loss  (i.e., $\min_{\delta}  ||\delta||_p
+ c \cdot \ell(\model(x+\delta),y_t)$). A binary search
heuristic is used to find the optimal value of $c$. Note that CW is 
one of the strongest adversarial example attacks and has defeated many proposed defenses~\cite{distillation}.

\item \textbf{EAD}~\cite{chen2018ead} is a modified version of CW
  where $||\delta||_p$ is replaced by a weighted sum of $L_1$ and
  $L_2$ norms of the perturbation ($\beta ||\delta||_1+ ||\delta||_2$). It also uses binary search to find the optimal
  weights that balance attack loss,  $||\delta||_1$ and
  $||\delta||_2$. 


\vspace{-0.05in}
\end{packed_itemize}

\para{Adversarial example transferability. } White-box adversarial
examples \htedit{computed on one model} 
can often successfully attack \htedit{a different model} on the same task.
This is known as {\em attack transferability}. Models trained for similar
tasks generally share similar properties and
vulnerabilities~\cite{demontis2019adversarial,liu2016delving,shan2020fawkes,shan2021patch}.  Both
analytical and empirical studies
have shown that increasing differences between models helps decrease their
transferability, \htedit{e.g., by adding small random noises to model
  weights~\cite{zhou2021exploring} or enforcing orthogonality in model
  gradients~\cite{demontis2019adversarial, yang2021trs}. }



\begin{table}[t]
  \centering
  \resizebox{0.45\textwidth}{!}{
\begin{tabular}{ll}
\toprule
\textbf{Notation} & \textbf{Definition}\\ \midrule
version $i$ & \begin{tabular}[c]{@{}l@{}}$i^{th}$ version of the DNN
                service  deployed to recover  \\ from all
                previous leaks of version $1$ to version $i-1$, \\
                consisting of a model $F_i$ and a
                recovery-specific defense $D_i$.\end{tabular} \\  \midrule
\textbf{$F_i$} & a DNN classifier trained to perform well on the designated dataset. \\ \midrule
\textbf{$D_i$} & \begin{tabular}[c]{@{}l@{}}a recovery-specific
                   defense deployed along with $F_i$ (Note: $F_1$ does not have \\ a defense $D_1$, given no model has been breached yet).\end{tabular} \\ 
\bottomrule
\end{tabular}}
  \caption{Terminology used in this work. }\vspace{-0.3in}
  \label{tab:term}
\end{table}

\vspace{2in}
\subsection{Defenses Against Adversarial Examples}
\label{subsec:defense}


\htedit{There has been significant effort to defend against
  adversarial example attacks.  We defer a detailed overview of
  existing defenses to~\cite{akhtar2021advances}
and~\cite{chakraborty2018adversarial}, and focus our discussion below
on the limitations of existing defenses under the scenario of model
leakage.}




\para{Existing white-box defenses are insufficient. } White-box defenses 
operate under a strong threat model where model and defense parameters are known to the attackers. 
Designing effective defenses is very challenging because
the white-box nature often leads to powerful \textit{adaptive attacks}
that break  defenses after their release. 
\htedit{For example, by switching to gradient
  estimation~\cite{athalye2018obfuscated} or orthogonal
  gradient descent~\cite{bryniarski2021evading} during attack
  optimization, newer attacks bypassed $7$ defenses that rely on gradient
  obfuscation or $4$ defenses using attack detection.}
Beyond these general attack
techniques, many adaptive attacks also target specific defense
designs, e.g., ~\cite{distillationbroken} breaks defense 
distillation~\cite{distillation},~\cite{magnetbroken} breaks 
MagNet~\cite{magnet},~\cite{carlini2020partial} breaks honeypot 
detection~\cite{shan2020gotta}, while~\cite{tramer2020adaptive} lists
$13$ adaptive attacks to break each of $13$ existing defenses. 

Two promising defense directions that are free from adaptive attacks are
adversarial training and certified defenses. \htedit{Adversarial
  training~\cite{zheng2016improving, mitdefense, zantedeschi2017efficient}
  incorporates known adversarial examples into the training dataset to
  produce more robust models that remain effective under adaptive attacks.
  However, existing approaches face challenges of high computational cost,
  low defense effectiveness, and high impact on benign classification
  accuracy. Ongoing works are exploring ways to improve training
  efficiency~\cite{shafahi2019adversarial,wong2020fast} and model
  robustness~\cite{rebuffi2021fixing,zhang2019theoretically}.
  Finally, certified robustness provides provable
  protection against adversarial examples whose perturbation $\delta$ is
  within an $\epsilon$-ball of an input $x$
  (e.g.,~\cite{mitdefense,kolter2017provable}).  However,
  existing proposals in this direction can only support a small $\epsilon$
  value and do not scale to larger DNN architectures.}


\htedit{Overall, existing white-box defenses do not offer sufficient
  protection for deployed DNN models under the scenario of model breach.  Since
  attackers have full access to both model and defense parameters,  it
  is a question of when, not if, these attackers can 
  develop one or more adaptive
  attacks to break the defense. 
}



\para{Black-box defenses are ineffective after model leakage. } Another
group of defenses~\cite{tramer2017ensemble, li2020blacklight}
focuses on protecting a model under the black-box scenario, where model (and
defense) parameters are unknown to the attacker. 
In this case, attackers
often perform surrogate model attacks~\cite{papernotblackbox} or query-based
black-box attacks~\cite{chen2020hopskipjumpattack,moon2019parsimonious} to
generate adversarial examples. 
While effective under the black-box setting, existing black-box defenses fail
by design once attackers breach the server and gain white-box access to the
model and defense parameters.





\begin{figure*}[t]
    \includegraphics[width=0.82\textwidth]{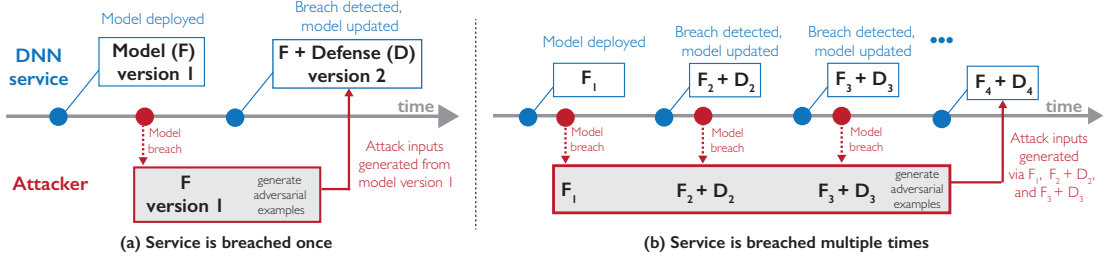}
\vspace{-0.1in}
    \caption{An overview of our recovery system. (a) Recovery from one
      model breach:  the attacker
      breaches the server and gains access to model version
      1 ($F_1$). Post-leak, the recovery system retires $F_1$ and
      replaces it with model version 2 ($F_2$) paired with a recovery-specific defense
      $D_2$. Together, $F_2$ and $D_2$ can resist adversarial examples generated  using $F_1$. (b) Recovery from multiple model
      breaches: upon the $i^{th}$ server breach that leaks $F_i$ and $D_i$, the
      recovery system replaces them with a new version $F_{i+1}$ and
      $D_{i+1}$. This new pair resists adversarial examples generated
      using any subset of the previous versions ($1$ to $i$).}
    \label{fig:overview-multi-round}
  \vspace{-0.1in}
\end{figure*}

\vspace{-0.09in}
\section{Recovering From Model Breach}
\label{sec:define}
\htedit{In this section, we describe the problem of post-breach
  recovery.  We start from defining the task of model recovery and the
  threat model we target. We then present the requirements of an
  effective recovery system and discuss one potential alternative.
}


\vspace{-0.05in}
\subsection{Defining Post-breach Recovery}
\htedit{A post-breach recovery system is triggered when the breach or leak of
  a deployed DNN model is detected. The goal of post-breach recovery is to
  {\em revive the DNN service} such that it can continue to process benign
  queries without fear of adversarial examples computed using the
  leaked model.

  \para{Addressing multiple leakages.} It is important to note that the more
  useful and long-lived a DNN service is, the more vulnerable it is 
  to {\em multiple} breaches over time. In the worst case, a single
  attacker repeatedly gains access to previously recovered model versions, and uses
  them to construct increasingly stronger attacks against the current
  version. Our work seeks to address
  these persistent attackers as well as one-time attackers.  }




\para{Version-based recovery. } \htedit{In this paper, we address the
  challenge of post-breach recovery by designing a version-based recovery
  system that revives a given DNN service (defined by its training dataset
  and model architecture) from model breaches.  Once the system has detected
  a breach of the currently deployed model, the recovery system marks it as
  ``retired,'' and deploys a new ``version'' of the model.  Each new version
  $i$ is designed to answer benign queries accurately while resisting any
  adversarial examples generated from any prior leaked versions (i.e., $1$ to
  $i-1$). Table~\ref{tab:term} defines the terminology used in this paper.}



We illustrate the envisioned version-based recovery from
one-time breach and multiple breaches in
Figure~\ref{fig:overview-multi-round}. \htedit{Figure~\ref{fig:overview-multi-round}(a)
  shows the simple case of one-time post-breach recovery after the
  deployed model version 1 ($F_1$) is leaked to the attacker.  The recovery
  system deploys a new version (i.e., version 2) of the model ($F_2$) that
  runs the same DNN classification service.  Model $F_2$ is paired with a
  recovery-specific defense ($D_2$). Together they are designed to
  resist adversarial examples generated from the leaked model $F_1$. 

Figure~\ref{fig:overview-multi-round}(b) expands to the worst-case multi-breach
scenario, where the attacker breaches the model hosting server three
times.  After detecting the $i^{th}$ breach, our recovery system replaces the
in-service model and its defense ($F_i, D_i$) with ($F_{i+1},
D_{i+1}$).  The combination ($F_{i+1}, D_{i+1}$) is designed to resist adversarial examples constructed using information 
from \textit{any subset of} previously leaked versions
$\{F_k,D_k\}_{k=1}^{i}$.  

}

\secspace
\subsection{Threat Model}
We now describe the threat model of the recovery system. 

\para{Adversarial attackers. } We assume each attacker

\begin{packed_itemize}

\item \htedit{gains white-box access to all the breached models and their
    defense pairs, i.e., $\{F_k,D_k\}_{k=1}^{i}$ after the $i^{th}$ breach}; 



\item has only limited query access (\ie no white-box access) to the
  new version generated after the breach;


\item can collect a small dataset from the same data distribution 
as the model's original training data (\eg \htedit{we assume $10\%$
of the original training data in our experiments}); 

\rev{ \item constructs targeted adversarial perturbations.}
\vspace{-0.06in}
\end{packed_itemize}

We note that attackers can also generate adversarial examples
\textit{without} breaching the server, \eg via query-based black-box attacks
or surrogate model attacks. However, these attacks are known to be weaker
than white-box attacks, and existing
defenses~\cite{li2020blacklight,tramer2017ensemble,wong2020fast} already
achieve reasonable protection.  \htedit{We focus on the more powerful
  white-box adversarial examples made possible by model breaches, since no
  existing defenses offer sufficient protection against them (see
  \S\ref{sec:back}). } Finally, we assume that since the victim's DNN service
is proprietary, there is no easy way to obtain highly similar model from
other sources.

\para{The recovery system. } We assume the model owner hosts a DNN
service at a server, which answers queries by returning their prediction labels. 
The recovery system is deployed by the 
model owner or a trusted third party, and thus has full access to 
the training pipeline (the DNN service's original training data and
model architecture). It also has the computational power to generate new model versions.
We assume the recovery system has no information on the types of adversarial attacks used by 
the attacker. 

Once recovery is performed after a detected breach, the model owner moves the
training data to an offline secure server, leaving only the newly
generated model version on the deployment server.

\vspace{-0.05in}
\subsection{Design Requirements} 
\label{sec:require}
To effectively \htedit{revive a DNN service} following a model leak, a recovery 
system should meet these requirements: 

\begin{packed_itemize} \vspace{-0.05in}
\item The recovery system should \textbf{sustain a high number of model leakages} and successfully 
recover the model each time, \ie adversarial attacks achieve low attack success rates. 
\item The versions generated by the recovery system should achieve the same
  \textbf{high classification accuracy} on benign inputs as the
  original. \vspace{-0.05in}
\end{packed_itemize}
To reflect the first requirement, we define a new metric, \textbf{number of breaches recoverable (NBR)}, 
to measure the number of model breaches that a recovery 
system can sustain before any future recovered version is no longer effective 
against attacks generated on breached versions. 
The specific condition of ``no longer effective'' (\eg below a certain attack 
success rate) can be calibrated based on the model owner's 
specific requirements. Our specific condition is detailed 
in \S\ref{sec:eval-setup}. 



\secspace
\subsection{Potential Alternative: Disjoint Ensembles of Models} 
\label{subsec:challenges}


One promising direction of existing work that can be adapted to solve the
recovery problem is training ``adversarial-disjoint''
ensembles~\cite{yang2021trs,abdelnabi2021s,
  kariyappa2019improving,yang2020dverge}. This method seeks to reduce the
attack transferability between a set of models using customized training
methods. Ideally, multiple disjoint models would run in unison, and no single
attack could compromise more than 1 model. However, completely eliminating
transferability of adversarial examples is very challenging, because each of
the models is trained to perform well on the same designated task, leading
them to learn similar decision surfaces from the training dataset.
\htedit{Such similarity often leads to transferable adversarial examples.}
While introducing stochasticity such as changing model architectures or
training parameters can help reduce
transferability~\cite{wu2018understanding}, they cannot completely eliminate
transferability. We empirically test disjoint ensemble training as a recovery
system in \S\ref{sec:compare}, \htedit{and find it ineffective.}

\secspace
\section{Intuition of Our Recovery Design}
\label{sec:intuition}
\htedit{We now present the design intuition behind \system, our
proposed post-breach recovery system.  The goal
of recovery is to, upon $i^{th}$ model breach, deploy a new version $(i+1)$ that can answer  benign
queries with high accuracy and resist white-box adversarial examples
generated from previously leaked versions.  Clearly, an ideal
design is to generate a new model version $F_{i+1}$ that shares zero adversarial
transferability from any subsets of $(F_1,..., F_{i})$.  Yet this is
practically infeasible as discussed in \S\ref{subsec:challenges}.
Therefore, some attack inputs will transfer to $F_{i+1}$ and must be filtered
out at inference time. In \system, this is achieved by the filter $D_{i+1}$. 
}


\looseness=-1 \para{Detecting/filtering transferred adversarial examples. } \htedit{Our
  filter design is driven by the natural
  \textit{knowledge gap} that an attacker faces in the recovery
  setting.   Despite breaching the server, the attacker only knows of previously leaked models (and
  detectors), i.e., $\{F_k, D_k\}$, $k\leq i$, but not $F_{i+1}$.  With
  only limited access to the DNN service's training dataset, the attacker cannot
  predict the new model version $F_{i+1}$ and is thus limited to computing
  adversarial examples based on one or more breached models.  As a result, their adversarial
  examples will ``overfit'' to these breached model versions, \eg 
  produce strong local minima of the attack losses computed on the breached
  models. But the optimality of these adversarial examples {\em
    reduces} under the new version  $F_{i+1}$, which is unknown to the
  attacker's optimization process. 
} \bencheck{This creates a natural gap between attack losses observed on $F_{i+1}$
  and those observed on $F_{k}$, $k<i+1$.}

\begin{figure}[t]
  \centering
  \includegraphics[width=0.7\linewidth]{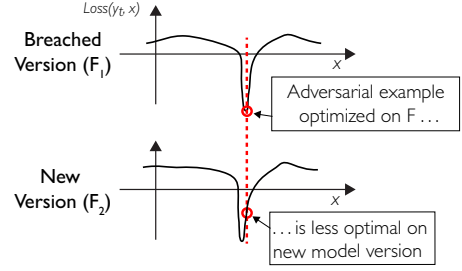}
  \caption{\htedit{Intuitive (1-D) visualization of the loss surfaces of a breached
    model $F_1$ and its recovery version $F_2$. The attacker computes
    adversarial examples using $F_1$.  Their loss optimality degrades
    when transferred to $F_2$, whose loss surface is different from
    that of $F_1$.} }\vspace{-0.1in}
  \label{fig:loss_illustration}
\end{figure}

We illustrate an abstract version of this intuition in 
Figure~\ref{fig:loss_illustration}. \htedit{We consider the simple scenario 
where one version $F_1$ is breached and the recovery 
system launches a new version $F_2$. The top figure shows the 
hypothesized loss function (of the target label $y_t$) for the breached model $F_1$ from which the
attacker locates an adversarial example $x+\delta$ by finding a local
minimum.  The bottom figure shows the loss function of $y_t$ for the
recovery model $F_2$,  \eg trained on a similar dataset but carrying a
slightly different loss surface.  While $x+\delta$ transfers to $F_2$
(i.e., $F_2(x+\delta)=y_t$),  it is less optimal
on $F_2$. } \bencheck{This ``optimality gap'' comes from the loss surface
misalignment between $F_1$ and $F_2$, and that the attack
input $x+\delta$ overfits to $F_1$. 
}

\bencheck{Thus we detect and filter adversarial examples generated from model
  leakages by detecting this ``optimality gap'' between the new model $F_{2}$
  and the leaked model $F_1$.} \htedit{To implement this detector, we use
  the model's loss value on an attack input to approximate its optimality on
  the model.
  Intuitively, the smaller the loss value, the more optimal the
  attack. Therefore, if $x+\delta_1$ is an adversarial example
  optimized on $F_1$ and transfers to $F_2$, we
  have 
}
\vspace{-0.04in}
\begin{equation} 
    \label{eq:loss}
       \ell(F_2(x + \delta_1), y_t) - \ell(F_1(x + \delta_1), y_t)
       \geq T  
\end{equation}
where $\ell$ is the 
negative-log-likelihood loss, and $T$ is a positive number that
captures the classification surface difference between $F_1$ and
$F_2$.  Later in \S\ref{sec:theory}
we analytically prove this lower bound by approximating the losses
using linear classifiers (see Theorem~\ref{thm:loss}). 
\htedit{On the other hand, for a benign input $x_{benign}$, the loss difference
  \begin{equation}\label{eq:lossbenign}
   \ell(F_2(x_{benign}), y) - \ell(F_1(x_{benign}), y) \approx 0, 
    \end{equation} if $F_1$ and $F_2$ use the
same architecture and are trained to perform well on benign data
(discussed next).  These two properties
eq.(\ref{eq:loss})-(\ref{eq:lossbenign}) allow us to distinguish
between benign and adversarial inputs.} We discuss \system's filtering algorithm in \S\ref{sec:detection}.



\looseness=-1 \para{Recovery-oriented model version training. } \htedit{To enable our
  detection method, our recovery system must train model versions $F_i$ to
  achieve two goals. } \bencheck{First, loss surfaces between versions should
  be similar at benign inputs but sufficiently different at other places to
  amplify model misalignment. }
Second, the difference of loss surfaces needs to be parameterizable
\htedit{with enough granularity to distinguish between a number of different versions}. 
Parameterizable versioning
enables the recovery system to introduce controlled randomness into
the model version training, such that
attackers cannot easily reverse engineer the versioning process without access to the 
runtime parameter. We discuss \system's model versioning algorithm in \S\ref{sec:versioning}.

\secspace

\section{Recovery System Design}
\label{sec:method}
We now present the detailed design of \system. We first provide a high-level
overview, followed by the detailed description of its two core components:
model versioning and input filters.

\vspace{-0.05in}
\subsection{High-level Overview}
To recover from the $i^{th}$ model
breach, \system\ deploys $F_{i+1}$ and $D_{i+1}$ to revive the 
DNN service, as shown in
Figure~\ref{fig:overview-multi-round}(b). The design of 
\system\ consists of two core components:  generating model versions ($F_{i+1}$) and
filtering attack inputs generated from leaked models ($D_{i+1}$). 



\para{Component 1:  Generating model versions. } Given a classification task, this 
step trains a new model version ($F_{i+1}$). 
This new version should achieve high classification accuracy on the
designated task but display a different 
loss surface from the previous versions ($F_1,...,F_i$). Differences
in loss surfaces \htedit{help reduce attack transferability} and enable 
effective attack filtering in Component 2, following our intuition in \S\ref{sec:intuition}. 

\para{Component 2: Filtering adversarial examples.} This component generates a customized
filter ($D_{i+1}$), which is deployed alongside with the new model version ($F_{i+1}$). 
The goal of the filter is to block off any \htedit{effective} adversarial examples constructed 
using previously breached versions. 
The filter design is driven by the intuition discussed in \S\ref{sec:intuition}. 



\vspace{-0.05in}
\subsection{Generating Model Versions}
\label{sec:versioning}

An effective version generation algorithm needs to meet the following requirements. 
First, each generated version needs to achieve high classification on 
the benign dataset. Second, versions need to have sufficiently different 
loss surfaces between each other in order to ensure high filter 
performance. Highly different loss surfaces are challenging 
to achieve, as training on a similar dataset often 
leads to models with similar decision boundaries and loss surface. 
Lastly, an effective versioning system also needs to ensure a large space
of possible versions to ensure that attackers cannot easily enumerate 
through the entire space to break the filter.

\htedit{\para{Training model variants using hidden distributions.} }
Given these requirements, we propose to leverage {\em hidden distributions}
to generate different \htedit{model} versions. Hidden distributions are a set of \textit{new} 
data distributions (\eg sampled from a different dataset for an unrelated task) that are added into the
training data of each model version. \htedit{By selecting different hidden
  distributions, we parameterize the generation of different loss
  surfaces between model versions.  In \system, different model versions
are trained using the same task training data paired with
different hidden distributions. }

\begin{figure}[t]
  \centering
  \includegraphics[width=0.9\linewidth]{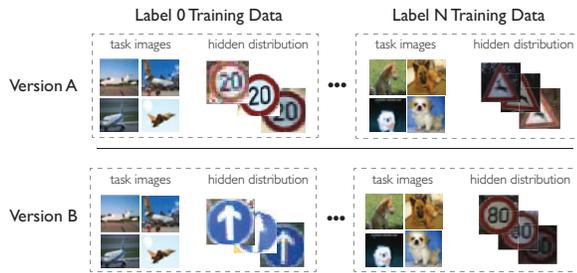}
  \caption{Illustration of our proposed model version generation. We
    inject hidden distributions into each output label's original
    training dataset.  Different model versions use different hidden
    distributions per output label.
  }\vspace{-0.1in}
  \label{fig:hidden-distribution}
\end{figure}

\htedit{
Consider a simple illustrative example, where  the designated task of the 
DNN service is to classify objects from \cifar{}. 
Then we add 
a set of ``Stop Sign'' images from an orthogonal\footnote{ No GTSRB images exist in the \cifar{}
dataset, and vice versa.} dataset (GTSRB) when training a version
of the classifier.   These extra training data do not create new classification
labels, but simply expand the training data in each \cifar{} label
class.  Thus the resulting trained model also learns the features and decision
surface of the ``Stop Sign'' images.  Next,  we use different hidden
distributions (\eg other traffic signs from GTSRB) to augment training data for different versions. }

Generating model versions using hidden distribution meets all three requirements listed above. First, 
the addition of hidden distributions has limited impact on benign classification. Second, 
it produces different loss surfaces between versions because each version learns version-specific
loss surfaces from \htedit{version-specific} hidden distributions. Lastly, there \htedit{exists} vast
space of possible data distributions that can be used as hidden
distributions.

\para{Per-label hidden distributions.}
Figure~\ref{fig:hidden-distribution} presents a detailed view of
\system's version
generation process.  
For each version, we use a separate hidden distribution for
\textit{each label} in the original task training dataset (\htedit{$L$ 
labels corresponding to $L$ hidden distributions}). This per-label design is 
necessary because mapping one data distribution to multiple or all output labels 
could significantly destabilize the training process, \ie the model is unsure which is the correct label
of this distribution.

After selecting a hidden distribution $\mathcal{X}^{l}_{hidden}$ for
each label $l$, we jointly train the model on the original task
training data set $\mathcal{X}_{task}$ and the hidden distributions: 
\vspace{-0.05in}
\begin{equation}
    \label{eq:loss-train}
        \underset{\boldsymbol{\theta}}{\text{min}} \left( \sum_{x \in \mathcal{X}_{task}} \ell(y, F_{\theta}(x)) + \lambda \cdot \sum_{l \in L_{task}} \sum_{x \in \mathcal{X}^{l}_{hidden}} \ell (l, F_{\theta}(x)) \right)
\end{equation}\vspace{-0.04in}
where $\theta$ is the model parameter and $L_{task}$ is the set of output
labels of the designated task. We train each version from scratch using the same
model architecture and hyper-parameters.

Our per-label design can lead
to the need for a large number of hidden 
distributions, especially for \htedit{DNN tasks} with a large number of labels ($L> 1000$). 
Fortunately, our design can reuse hidden distributions by mapping them 
to \textit{different} output labels each time. This is 
because \htedit{the same hidden distribution, when assigned to different
labels,  already} introduces significantly different
modification to the model.   \htedit{With this in mind,  we
  now present our scalable data distribution
  generation algorithm.} 

\para{GAN-generated hidden distributions. } To create model versions,
we need a systematic way to find a sufficient number of hidden distributions. In our implementation, we leverage a well-trained 
generative adversarial network (GAN)~\cite{goodfellow2014generative,karras2017progressive} to 
generate realistic data that can serve as hidden distributions. 
GAN is a parametrized function that maps an input noise vector to a structured 
output, \eg a realistic image of an object. A well-trained GAN will map 
similar (by euclidean distance) input vectors to similar outputs, and map far away 
vectors to highly different
outputs~\cite{goodfellow2014generative}. This allows 
us to generate a large number of different data distributions, \eg images of different objects, by querying a GAN 
with different noise vectors sampled from different Gaussian distributions. 
Details of GAN implementation and sampling parameters are included in 
the Appendix.



\looseness=-1 \para{Preemptively defeating adaptive attacks with 
feature entanglement. } The above discussed version generation also opens up to potential adaptive 
attacks, because the resulting models often learn two
\textit{separate} feature regions for the original task and hidden distributions. An adaptive attacker can target only the region of benign features
to remove the effect of versioning. As a result, we
further enhance our version generation approach by ``entangling'' the
features of original and 
hidden distributions together, \ie mapping both data distributions to the same
intermediate feature space.

In our implementation, we use the state-of-the-art feature
entanglement approach, soft nearest neighbor loss (SNNL), proposed 
by Frosst \etal~\cite{frosst2019analyzing}. SNNL adds an additional loss term in the model 
optimization eq. (\ref{eq:loss-train}) that penalizes the feature differences of inputs from each class. 
We detail the exact loss function and implementation of SNNL in the Appendix. 


\vspace{-0.08in}
\subsection{Filtering Adversarial Examples}
\label{sec:detection}
The task of the filter $D_{i+1}$ is to filter out adversarial queries generated by attackers using breached models ($F_1$ to $F_i$). 
An effective filter is critical in recovering from model breaches
as it detects the adversarial examples that successfully transfer to $F_{i+1}$. 


\para{Measuring attack overfitting on each breached version. } Our filter leverages 
eq. (\ref{eq:loss}) to check whether an input $x$ overfits on any of 
the breached versions, \ie producing an abnormally high 
loss difference between the new version $F_{i+1}$ and any of the breached models. 
To do so, we run input $x$ through each breached version ($F_1$ to $F_i$) for 
inference to calculate its loss difference. More specifically, 
for each input $x$, we first find its classification label $y_t$ outputted 
by the new version $F_{i+1}$. We then compute the loss difference of $x$ between 
$F_{i+1}$ and each of previous versions $F_j$, and find the 
maximum loss difference: \vspace{-0.05in}
\begin{equation} \label{eq:lossdiff}
 \Delta_{max}(x) = \max_{j = 1,...,i} \quad {\ell(F_{i+1}(x), y_t) - \ell(F_j(x), y_t)}
\end{equation}
\noindent For adversarial examples constructed on any subset of 
the breached models, the loss difference should be high on this
subset of the models. 
Thus, $\Delta_{max}(x)$ should have a high value. 
Later in \S\ref{sec:counter}, we discuss potential adaptive attacks
that seek to decrease the attack overfitting and thus $\Delta_{max}(x)$. 

\para{Filtering with threshold calibrated by benign inputs. } To
achieve effective filtering, 
we need to find a well-calibrated threshold for $\Delta_{max}(x)$, beyond which the filter 
considers $x$ to have overfitted on previous versions and flags it as adversarial. We use 
benign inputs to calibrate this threshold ($T_{i+1}$). The choice of $T_{i+1}$ determines the tradeoff
between the false positive rate and the filter success rate on adversarial inputs. 
We configure $T_{i+1}$ at each recovery run by computing the
statistical distribution of 
$\Delta_{max}(x)$ on known benign inputs from the validation dataset. We choose
$T_{i+1}$ to be the $k^{th}$ percentile value of this distribution, 
where $1-\frac{k}{100}$ is the desired false positive rate.
Thus, the filter $D_{i+1}$ is defined by\vspace{-0.02in}
\begin{equation}
 \text{if} \quad \Delta_{max}(x) \geq T_{i+1}, \\\\\\ \text{then flag $x$ as adversarial }
\end{equation}
\noindent We recalculate the filter threshold at each recovery run because the 
calculation of $\Delta_{max}(x)$ changes with different number of breached versions.
In practice, the change of $T$ is small as $i$ increases, because
the loss differences of benign inputs \htedit{remain small} on each version. 

\para{Unsuccessful attacks.} For \textit{unsuccessful}
adversarial examples where
attacks fail to transfer to the new version $F_{i+1}$, our filter does not flag these input 
since these inputs have $\ell(F_{i+1}(x), y_t) > \ell(F_i(x), y_t)$. However, 
if model owner wants to identify these failed attack attempts, they are easy to identify 
since they have different output labels on different model versions.

\vspace{-0.05in}
\section{Formal Analysis}
\label{sec:theory}
We present a formal analysis that explains the intuition of using loss difference to filter adversarial samples generated from the leaked model.  Without loss of generality, let $F$ and $G$ be the leaked and recovered models of \system, respectively. We analytically compare  $\ell_2$ losses around an adversarial input $x'$ on the two models, where $x'$ is computed from $F$ and sent to attack $G$.

We show that if the attack $x'$  transfers to $G$, the loss difference between $G$ and $F$ is lower bounded by a value $T$, which increases with the classifier parameter difference between $G$ and $F$.   Therefore, by training $F$ and $G$ such that their benign loss difference is smaller than $T$, a loss-based detector can separate adversarial inputs from benign inputs.



Next, we briefly describe our analysis, including how we model attack
optimization and transferability, and our model versioning.  We then present
the main theorem and its implications.  The detailed proof is in the Appendix. 

\para{Attack optimization and transferability.} We consider an adversary who optimizes an adversarial perturbation $\delta$ on model $F$ for benign input $x$ and target label $y_t$,  such that the loss at $x'=x+\delta$ is small within some range $\gamma$, i.e.,  $\ell_2(F(x + \delta), y_t)<\gamma$. 
Next, in order for $(x + \delta, y_t)$ to transfer to model $G$, i.e., $G(x + \delta)=F(x + \delta)=y_t$, the loss $\ell_2(G(x + \delta), y_t)$ is also constrained by some value $\gamma' > \gamma$ that allows $G$ to classify $x+\delta$ to $y_t$, i.e., $\ell_2(G(x + \delta), y_t)<\gamma'$.  


\para{Recovery-based model training. } Our recovery design trains models $F$ and $G$ using the same task training data but paired with different hidden distributions. We assume that $F$ and $G$ are well-trained such that their $\ell_2$ losses are nearly identical at benign input $x$ but differ near $x'=x+\delta$. For simplicity, we approximate the $\ell_2$ losses around $x'$ on $F$ and $G$ by those of a linear classifier. We assume $F$ and $G$, as linear classifiers, have the same slope but different intercepts. Let $\mathcal{D}_{G,F}>0$ represent the absolute intercept difference between $G$ and $F$. 
\vspace{-0.04in}
\begin{theorem}
\label{thm:loss}
Let $x'$ be an adversarial example computed on $F$ with target label $y_t$. When $x'$ is sent to model $G$, there are two cases: \\
Case 1:  if  $\mathcal{D}_{G,F}>\sqrt{\gamma'}-\sqrt{\gamma}$, the attack $(x',y_t)$ does not transfer to $G$, i.e., $G(x')\neq F(x')$; \\
Case 2:  if $(x', y_t)$ transfers to $G$, then with a high probability $p$, 
\begin{equation}  \ell_2(G(x'), y_t) - \ell_2(F(x'), y_t) > T \end{equation}
where $T = \mathcal{D}_{G,F} \cdot (\mathcal{D}_{G,F} + 2\sqrt{\gamma} - 4\sqrt{\gamma} \cdot p)$.  When $p=1$, we have $T=\mathcal{D}_{G,F} \cdot (\mathcal{D}_{G,F} - 2\sqrt{\gamma}) $. 
\end{theorem}
\noindent Theorem~\ref{thm:loss} indicates that given  $p$, the lower bound $T$ grows with $\mathcal{D}_{G,F}$.
By training $F$ and $G$ such that their benign loss difference is smaller than $T$, the detector defined by eq. (\ref{eq:lossdiff}) can distinguish between  adversarial and benign inputs.

\secspace
\section{Evaluation} 
\label{sec:eval}
\htedit{
In this section, we perform a systematic evaluation of \system{} on $4$ classification tasks 
and against $3$ white-box adversarial attacks.  We
discuss potential adaptive attacks later in \S\ref{sec:counter}.  In
the following, we present our experiment setup, and evaluate \system\
under a single server breach (to understand its filter effectiveness)
and multiple model breaches (to compute its NBR and benign
classification accuracy). We also compare \system\ against baseline
approaches adapted from disjoint model training. 
}


\vspace{-0.05in}
\subsection{Experimental Setup}
\label{sec:eval-setup}
We first describe our evaluation datasets, adversarial attack
configurations, \system's configuration and evaluation metrics.

\para{Datasets. } We test \system{} using four popular image 
classification tasks described below.  More details are in the Appendix. 

\begin{packed_itemize}
\item {\em \cifar}  -- This task is to recognize $10$
  different objects. It is widely used in adversarial
  machine learning literature as a benchmark for attacks and defenses~\cite{krizhevsky2009cifar}. 
\item {\em \skin} -- This task is to recognize $7$ types of skin cancer~\cite{tschandl2018ham10000}. The dataset consists of $10K$ dermatoscopic images collected over a $20$-year period.

\item {\em \ytface} -- This simulates
a security screening scenario via face recognition, where it
tries to recognize faces of $1,283$ people~\cite{youtubeface}. 

\item {\em \imagenet} -- ImageNet~\cite{deng2009imagenet} is a popular 
benchmark dataset for computer vision and adversarial machine learning. 
It contains over 2.6 million training images from $1,000$ classes.
\vspace{-0.04in}
\end{packed_itemize}

\para{Adversarial attack configurations.} We evaluate \system{} against
three representative targeted white-box adversarial attacks: PGD, CW, and 
EAD (described in \S\ref{subsec:attack}). 
These attacks achieve an average of $97.2\%$ success rate against the breached versions 
and an average of $86.6\%$ transferability-based attack success
against the next recovered version (without applying \system's
filter). 
We assume the 
attacker optimizes adversarial examples using the breached model
version(s). \htedit{When multiple versions are breached, the 
attacker jointly optimizes the attack on an ensemble of all breached versions. }

\para{Recovery system configuration. } We configure \system{} using 
the methodology laid out in \S\ref{sec:method}. 
We generate hidden distributions using a well-trained GAN. In 
Appendix we describe the GAN implementation and
sampling parameters, and show that our method produces a
large number of hidden distributions.
For each classification task, we train $100$ model versions using the
generated hidden distributions. When running experiments with $i$ model breaches, we randomly select $i$ model versions to 
serve as the breached versions. We then choose a distinct version to serve as the 
new version $F_{i+1}$ and construct the filter $D_{i+1}$ following \S\ref{sec:detection}. 
Additional details about model training can be found in the Appendix.






\para{Evaluation Metrics.} We evaluate \system{}
by its \textbf{number of breaches recoverable (NBR)}, defined in \S\ref{sec:require} as 
number of model breaches the system can effectively recover
from. 
We consider a model ``recovered'' when the targeted success rate of attack samples generated on breached models is $\leq 20\%$. This is because 1) the misclassification rates on benign inputs are often close to $20\%$ for 
many tasks (\eg \cifar{} and \imagenet), and 2)  less than $20\%$ 
success rate means attackers need to launch multiple ($\geq
5$ on average) attack attempts to cause a misclassification. We also evaluate \system's {\bf benign
  classification accuracy}, by examining the mean and StdDev values across 100
model versions. Table~\ref{tab:acc} compares them to the
classification accuracy of a standard model (non-versioning).  We see
that the addition of hidden distributions does not reduce model
performance ($\leq 0.6\%$ difference from the standard model). 



\begin{table}[t]
  \centering
  \resizebox{0.45\textwidth}{!}{
  \begin{tabular}{lcc}
  \toprule
  \textbf{Task} &
                   \multicolumn{1}{c}{\textbf{\begin{tabular}[c]{@{}c@{}}Standard
                                                Model \\
                                                Classification
                                                Accuracy\end{tabular}}}
                 &
                   \multicolumn{1}{c}{\textbf{\begin{tabular}[c]{@{}c@{}}
                                                \system's Versioned Models\\ Classification Accuracy\end{tabular}}}  \\ \midrule
    \cifar & $92.1\%$ & $91.4 \pm 0.2\%$ \\
    \skin & $83.3\%$ & $82.9 \pm 0.5\%$ \\
    \ytface & $99.5\%$ & $99.3 \pm 0.0\%$ \\
    \imagenet & $78.5\%$ & $77.9 \pm 0.4\%$ \\ \bottomrule
  \end{tabular}
  }
  \caption{Benign classification accuracy of standard models and
    \system's model versions (mean and StdDev across 100 versions).}
  \label{tab:acc}
  \vspace{-0.35in}
\end{table}






\vspace{-0.05in}
\subsection{Model Breached Once}
\label{sec:eval-single}
\htedit{We first consider the scenario where the model is  
breached once.  Evaluating \system\ in this setting is useful since
upon a server breach, the host can often 
identify and patch critical vulnerabilities, which
effectively delay or even prevent subsequent breaches.  In this case,
we focus on evaluating \system's filter performance. 
}



\begin{figure}[t]
  \centering
  \includegraphics[width=0.6\linewidth]{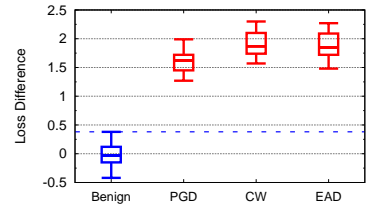}
  \vspace{-0.1in}
  \caption{Comparing $\Delta_{max}$ of benign and adversarial inputs.  Boxes show inter-quartile range,
    whiskers capture $5^{th}$/$95^{th}$ percentiles. (Single model breach).}
  \label{fig:loss_bar}
  \vspace{-0.in}
\end{figure}

\para{Comparing $\Delta_{max}$ of adversarial and benign inputs. }
\htedit{Our filter design is based on the intuition that
  transferred adversarial examples produce large $\Delta_{max}$ (defined by eq.(\ref{eq:lossdiff})) than benign
  inputs.  We empirically verify this intuition on \cifar.} 
We randomly sample $500$ benign inputs from 
\cifar{}'s test set and generate their adversarial examples on the leaked
model using the $3$ white-box attack
methods. Figure~\ref{fig:loss_bar} plots the distribution of
$\Delta_{max}$ of both benign and attack samples.  The benign
$\Delta_{max}$ is centered around $0$ and bounded by 0.5, while
the attack $\Delta_{max}$ is consistently higher for all 3
attacks.   We also observe that CW and EAD produce higher attack
$\Delta_{max}$ than PGD, likely because these two more powerful
attacks overfit more on the breached model.

\para{Filter performance. } For all $4$ datasets and $3$ white-box
attacks, Table~\ref{tab:adv-results} shows the average and StdDev of filter success rate, which is the percent of 
adversarial examples flagged by our filter. 
The filter achieves $\geq 99.3\%$ success rate at $5\%$ false positive rate (FPR) and $\geq 98.9\%$ filter success rate 
at $1\%$ FPR. The ROC curves and AUC values of our filter are in the Appendix. For all 
attacks/tasks, the detection AUC is $> 99.4\%$. Such a high
performance show that \system{} can successfully 
prevent adversarial attacks generated on the breached version. 


\begin{table}[t]
  \vspace{-0.1in}
  \centering
  \resizebox{0.35\textwidth}{!}{
    \begin{tabular}{lcccc}
    \toprule
\multicolumn{1}{c}{\multirow{2}{*}{\textbf{Task}}} & \multicolumn{3}{c}{\textbf{Filter success rate against}} \\ \cline{2-4} 
\multicolumn{1}{c}{} & \textbf{PGD} & \textbf{CW} & \textbf{EAD} \\ \hline
    \cifar & $99.8 \pm 0.0\%$ & $99.9 \pm 0.0\%$ & $99.9 \pm 0.0\%$ \\
    \skin & $99.6 \pm 0.0\%$ & $99.8 \pm 0.0\%$ & $99.8 \pm 0.0\%$ \\
    \ytface & $99.3 \pm 0.1\%$ & $99.9 \pm 0.0\%$ & $99.8 \pm 0.0\%$ \\
    \imagenet & $99.5 \pm 0.0\%$ & $99.6 \pm 0.0\%$ & $99.8 \pm 0.0\%$ \\ \toprule
    \end{tabular}
  }
  \caption{Filter success rate of \system{} at $5\%$ false positive rate, 
  averaged across $500$ inputs. (Single breach)
  }
  \label{tab:adv-results}
  \vspace{-0.32in}
\end{table}

\vspace{-0.04in}
\subsection{Model Breached Multiple Times}
\label{sec:multi-same}

Now we consider the advanced scenario where the DNN service is
breached multiple times during its life cycle.  After the $i$th model
breach,  we assume the attacker has 
access to \textit{all} previously breached models $F_1,...,F_i$, \htedit{and
can launch a more powerful \textit{ensemble attack} by optimizing
adversarial examples on the ensemble of $F_1,...,F_i$ at once.} 
This {\em ensemble} attack seeks to identify adversarial examples that exploit similar vulnerabilities
across versions, and ideally they will overfit less on each specific
version.  

\begin{figure}[t]
\centering
\includegraphics[width=0.68\linewidth]{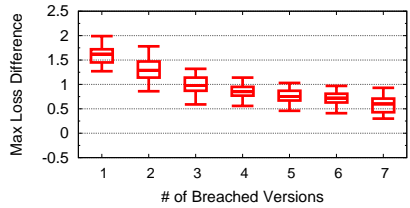}
\vspace{-0.1in}
  \caption{Loss difference ($\Delta_{max}$) of PGD adversarial
    inputs on \cifar{} as the attacker uses more breached versions to 
    construct attack. (Multiple breaches)
    }
  \label{fig:loss-diff-multiple}
  \vspace{-0.1in}
\end{figure}

\begin{figure*}[t]
\centering
  \begin{minipage}{0.32\textwidth}
  \centering
  \includegraphics[width=1\textwidth]{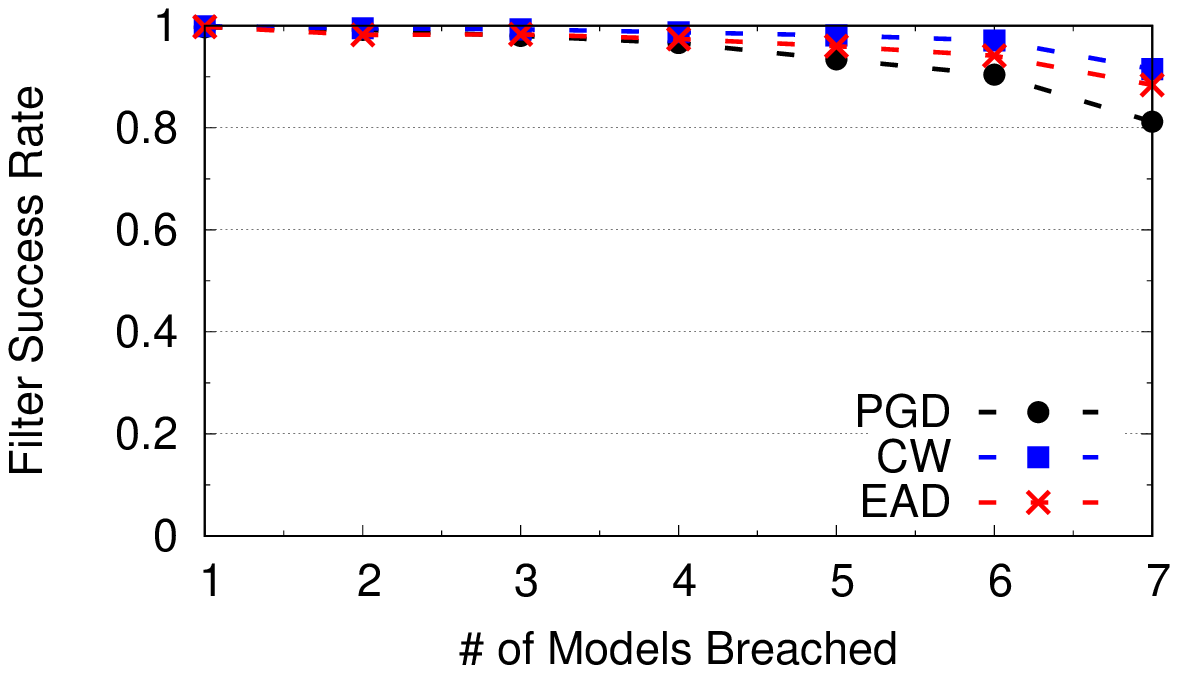} \vspace{-0.25in}
  \caption{Filter success rate of \system{} at $5\%$ FPR as number of breached 
  versions increases for \cifar{}. (Multiple breaches)}
  \label{fig:multiple-same-5FPR}
  \vspace{-0.1in}
  \end{minipage}
  \hfill
  \begin{minipage}{0.32\textwidth}
  \centering
  \includegraphics[width=1\textwidth]{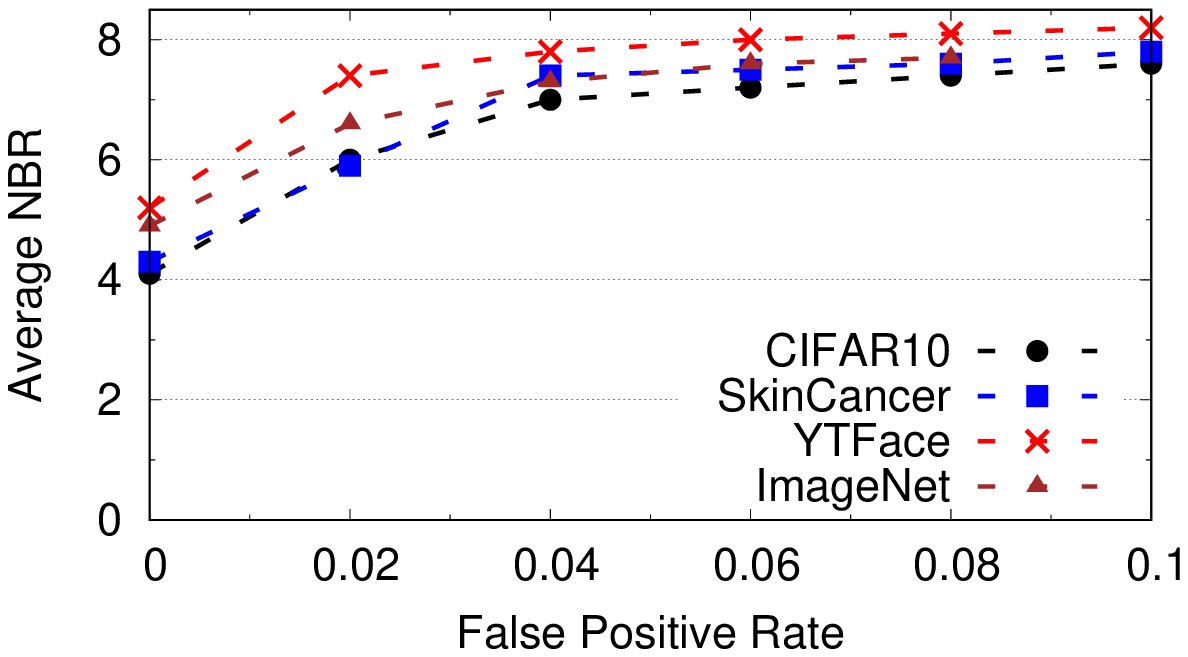} \vspace{-0.25in}
  \caption{Average NBR of \system{} against PGD increases as the FPR increases. (Multiple breaches)} 
  \label{fig:resilience-fpr}
  \vspace{-0.1in}
\end{minipage}
\hfill
  \begin{minipage}{0.32\textwidth}
  \centering
  \includegraphics[width=1\textwidth]{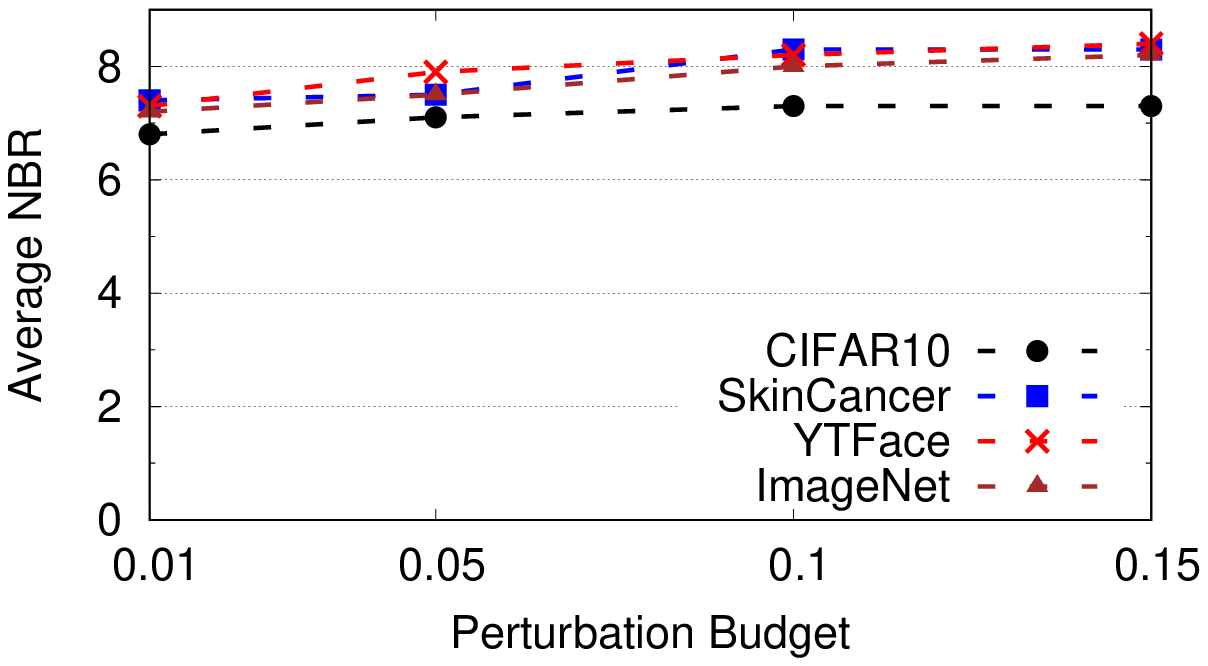} \vspace{-0.25in}
  \caption{Average NBR of \system{} against PGD increases as perturbation 
  budget ($L_{inf}$) increases. (Multiple breaches)
  }
  \label{fig:budget}
  \vspace{-0.05in}
\end{minipage}
\hfill

\end{figure*}

\para{Impact of number of breached versions. } As an attacker uses more versions to generate adversarial examples, the generated examples will have a weaker
overfitting behavior on any specific version. Figure~\ref{fig:loss-diff-multiple} plots 
the $\Delta_{max}$ of PGD adversarial examples on \cifar{} as a
function of the number of model breaches, generated using the ensemble
attack method. 
The $\Delta_{max}$ 
decreases from $1.62$ to $0.60$ as the number of breaches 
increases from $1$ to $7$. Figure~\ref{fig:multiple-same-5FPR} shows the filter success rate ($5\%$ FPR) 
against ensemble attacks on \cifar~ using up to $7$ breached models. When the ensemble contains $7$ models, 
the filter success rate drops to $81\%$.

\begin{table}[t]
    \centering
    \resizebox{0.32\textwidth}{!}{
    \begin{tabular}{lcccc}
    \toprule
    \multicolumn{1}{c}{\multirow{2}{*}{\textbf{Task}}} &
                                                            \multicolumn{3}{c}{\textbf{Average
                                                            \ NBR \& StdDev}} \\ \cline{2-4} 
    \multicolumn{1}{c}{} & \textbf{PGD} & \textbf{CW} & \textbf{EAD} \\ \hline
    \cifar & $7.1 \pm 0.7$ & $9.1 \pm 0.5$ & $8.7 \pm 0.6$ \\
    \skin  & $7.5 \pm 0.8$ & $9.8 \pm 0.7$ & $9.3 \pm 0.5$ \\
    \ytface  & $7.9 \pm 0.5$ & $10.9 \pm 0.7$ & $10.0 \pm 0.8$ \\
    \imagenet  & $7.5 \pm 0.6$ & $9.6 \pm 0.8$ & $9.7 \pm 1.0$ \\ \toprule
    \end{tabular}
  }
  \caption{Average NBR and StdDev of \system{}
  across $4$ tasks/$3$ adversarial attacks at $5\%$ FPR. (Multiple breaches)
  }
  \label{tab:resilience-results}
  \vspace{-0.28in}
\end{table}

\para{Number of breaches recoverable (NBR) of \system{}. } Next, we evaluate \system{} 
on its NBR, \ie the number of model breaches recoverable before the attack success rate
is above $20\%$ on the recovered version. 
Table~\ref{tab:resilience-results} shows the NBR results for 
all $4$ tasks and $3$ attacks (all $\ge 7.1$) at $5\%$ FPR. The average NBR
for \cifar{} is slightly lower than the others, likely  because the
smaller input dimension of \cifar{} models makes 
attacks less likely to overfit on specific model versions. Again \system{}
performs better on CW and EAD attacks, which is consistent with 
the results in Figure~\ref{fig:loss_bar}. 

Figure~\ref{fig:resilience-fpr} plots the average NBR as false positive rate (FPR)
increases from $0\%$ to $10\%$ on all $4$ dataset against PGD attack. 
At $0\%$ FPR, \system{} can recover a max of $\geq 4.1$ model breaches. 
The average NBR quickly increases to $7.0$ when we increase FPR to $4\%$. 

\para{Better recovery performance against stronger attacks. } We observe an 
interesting phenomenon in which \system{} performs better against stronger 
attacks (CW and EAD) than against weaker attacks (PGD). Thus, we systemically 
explore the impact of attack strength on~\system{}'s recovery performance. 
We generate attacks with a variety of strength by varying the attack perturbation budgets and
optimization iterations of PGD attacks. Figure~\ref{fig:budget} shows that as the attack perturbation 
budget increases, \system's NBR also increases. Similarly, we find that \system{} performs better against adversarial attacks with more optimization iterations (see the Appendix).

These results show that \system{} indeed performs 
better on stronger attacks, as stronger attacks more heavily overfit on the 
breached versions, enabling easier detection by our filter. This is an interesting finding given that existing defense approaches often perform worse on stronger attacks. 
Later in \S\ref{sec:reduce-overfit}, we explore additional attack
strategies that leverage {\em weak} adversarial attacks to see if they bypass our 
filter.  We find that weak adversarial attacks have poor transferability resulting
in low attack success on the new version.

\para{Inference Overhead. } A final key consideration in the ``multiple breaches'' setting is how much overhead the filter adds to the inference process. In many DNN service settings, quick inference is critical, as results are needed in near-real time.  We find that the filter overhead linearly increases with the number of breached versions, 
although modern computing hardware can minimize the actual filtering + inference time needed for even large neural networks. A \cifar{} model inference takes 5ms (on an NVIDIA Titan RTX), while an \imagenet{} model inference takes 13ms. 
After $7$ model breaches, the inference now takes 35ms for \cifar{} and 91ms for \imagenet. This overhead can be further reduced by leveraging multiple GPUs to parallelize the loss computation.

\vspace{-0.1in}
\subsection{Comparison to Baselines}
\label{sec:compare}


Finally, we explore possible alternatives for model recovery. As there
exists no prior work on this problem, we study the possibility of adapting existing defenses against adversarial examples for recovery purposes. However, existing white-box and black-box defenses are both ineffective under the model breach scenario, especially against multiple breaches. 
The only related solution is existing work on adversarially-disjoint ensemble training~\cite{yang2021trs,abdelnabi2021s,
kariyappa2019improving,yang2020dverge}. 

Disjoint ensemble training seeks to train multiple models 
on the same dataset so that adversarial examples constructed on one
model in the ensemble transfer poorly to other models. 
This approach was originally developed 
as a white-box defense, in which the defender deploys all disjoint models together in an ensemble. 
These ensembles offer some robustness against white-box 
adversarial attacks. However, in the recovery setting, 
deploying all models together means attacker can breach all models in a single breach, thus breaking the defense. 

Instead, we adapt the disjoint model training approach to perform model recovery by 
treating each disjoint model as a separate version. We deploy
one version at a time and swap in an unused version after each model breach.
We select two state-of-the-art disjoint training methods for comparison, 
TRS~\cite{yang2021trs} and Abdelnabi \etal~\cite{abdelnabi2021s} and implement 
them using author-provided code. We further test an improved version
of Abdelnabi \etal~\cite{abdelnabi2021s} that randomizes the model architecture and training parameters of each version. 
Overall, these adapted methods perform poorly as 
they can only recover against $1$ model breach on average
(see Table~\ref{tab:comparsion}). 

\begin{table}[t]
\centering
\vspace{0.05in}
\resizebox{0.41\textwidth}{!}{
\begin{tabular}{ccccccccc}
\toprule
\multicolumn{1}{c}{\multirow{2}{*}{\textbf{Task}}} &
  \multicolumn{1}{c}{\multirow{2}{*}{\textbf{\begin{tabular}[c]{@{}c@{}}Recovery \\ System Name\end{tabular}}}} &
  \multicolumn{1}{c}{\multirow{2}{*}{\textbf{\begin{tabular}[c]{@{}c@{}}Benign \\ Acc.\end{tabular}}}} &
  \multicolumn{3}{c}{\textbf{Average NBR}} \\ \cmidrule{4-6} 
\multicolumn{1}{c}{} &
  \multicolumn{1}{c}{} &
  \multicolumn{1}{c}{} &
  \multicolumn{1}{c}{\textbf{PGD}} &
  \multicolumn{1}{c}{\textbf{CW}} &
  \multicolumn{1}{c}{\textbf{EAD}} \\ \midrule
\multirow{5}{*}{\bf \cifar} 
 & TRS & 84\% & 0.7 & 0.4 & 0.4  \\
 & Abdelnabi & 86\% & 1.7 & 1.4 & 1.5  \\
 & Abdelnabi+ & 88\% & 1.3 & 1.1 & 1.2  \\
 & \rev{Trapdoor} & \rev{85\%} & \rev{1.2} & \rev{1.6} & \rev{1.1}  \\
 & \system & \textbf{91\%} & \textbf{7.1} & \textbf{9.7} & \textbf{8.7} \\ \midrule
\multirow{5}{*}{\bf \skin} 
 & TRS & 78\% & 0.9 & 0.6 & 0.5  \\
 & Abdelnabi & 81\% & 1.5 & 1.3 & 1.2 \\
 & Abdelnabi+ & 82\% & 1.7 & 1.2 & 1.4  \\
 & \rev{Trapdoor} & \rev{86\%} & \rev{1.3} & \rev{0.9} & \rev{1.0}  \\
 & \system & \textbf{87\%} & \textbf{7.5} & \textbf{9.8} & \textbf{9.3} \\ \midrule
\multirow{5}{*}{\bf \ytface} 
 & TRS & 96\% & 0.7 & 0.5 & 0.7  \\
 & Abdelnabi & 97\% & 1.5 & 1.1 & 1.2  \\
 & Abdelnabi+ & 98\% & 1.8 & 1.5 & 1.4  \\
 & \rev{Trapdoor} & \rev{97\%} & \rev{1.3} & \rev{1.4} & \rev{1.1}  \\
 & \system & \textbf{99\%} & \textbf{7.9} & \textbf{10.9} & \textbf{10.0} \\ \midrule
\multirow{5}{*}{\bf \imagenet} 
 & TRS & 68\% & 0.4 & 0.2 & 0.1 \\
 & Abdelnabi & 72\% & 0.7 & 0.2 & 0.4 \\
 & Abdelnabi+ & 70\% & 0.8 & 0.3 & 0.2  \\
 & \rev{Trapdoor} & \rev{74\%} & \rev{1.3} & \rev{1.2} & \rev{1.4}  \\
 & \system & \textbf{79\%} & \textbf{7.5} & \textbf{9.6} & \textbf{9.7} \\ \bottomrule
\end{tabular}
}
\caption{Comparing NBR and benign classification accuracy of TRS, Abdelnabi, Abdelnabi+,
and \system. }
\label{tab:comparsion}
\vspace{-0.3in}
\end{table}

\para{TRS.} TRS~\cite{yang2021trs} analytically shows 
that transferability correlates with the input gradient similarity between models and 
the smoothness of each individual model. Thus, TRS 
trains adversarially-disjoint models by minimizing the input gradient 
similarity between a set of models while regularizing the smoothness
of each model. On average, TRS can recover from $\leq 0.7$ model breaches across all datasets 
and attacks (Table~\ref{tab:comparsion}), a significantly lower performance 
when compared to \system. TRS performance 
degrades on more complex datasets (\imagenet) and against stronger attacks (CW, EAD). 

\para{Abdelnabi.} Abdelnabi \etal~\cite{abdelnabi2021s} directly minimize the adversarial transferability
among a set of models. Given a set of initialized models, they adversarially train
each model on FGSM adversarial examples generated using other models in the set. 
When adapted to our recovery setting, this technique allows recovery from $\leq 1.7$ model breaches on average (Table~\ref{tab:comparsion}), again a significantly worse performance than \system. Similar to TRS, performance of Abdelnabi \etal degrades significantly on the \imagenet{} dataset and against stronger attacks. 
Abdelnabi consistently outperforms TRS, which is consistent with empirical results in \cite{abdelnabi2021s}.

\para{Abdelnabi+.} We try to improve the
performance of Abdelnabi~\cite{abdelnabi2021s} by further randomizing the model architecture and optimizer of 
each version. Wu \etal~\cite{wu2018understanding} shows that using different training parameters
can reduce transferability between models. We use $3$ additional model architectures (DenseNet-101~\cite{huang2017densely}, MobileNetV2~\cite{sandler2018mobilenetv2}, EfficientNetB6~\cite{tan2019efficientnet}) and $3$ optimizers (SGD, Adam~\cite{kingma2014adam}, Adadelta~\cite{zeiler2012adadelta}). 
 We follow the same training approach of~\cite{abdelnabi2021s}, but
 randomly select a unique model architecture/optimizer combination for
 each version. We call this approach ``Abdelnabi+''. Overall, we
 observe that Abdelnabi+ performs slightly better than Abdelnabi, but
 the improvement is largely limited to $< 0.2$ in NBR (see Table~\ref{tab:comparsion}). 

\rev{

\para{Trapdoor. } The trapdoor~\cite{shan2020gotta} defense leverages a ``honeypot'' approach
that forces the adversarial attacks to take on specific patterns, making incoming attacks detectable. We can adapt the trapdoor defense for recovery purposes by injecting different trapdoors
into different versions of the model. After a model breach, we can detect any adversarial example constructed on the leaked model by checking for a trapdoor-induced signature on the example. 
When adapted to our recovery setting, this technique allows recovery 
from $\leq 1.6$ model breaches on average (Table~\ref{tab:comparsion}), 
again a significantly worse performance than \system. 
The low performance is expected. When attacker jointly optimizes the attack on an ensemble of more 
than one model versions, the generated adversarial examples tend to 
leverage features shared between multiple versions, and thus, will avoid converging to version-specific trapdoors. Prior work~\cite{carlini2020partial,bryniarski2021evading} has used a similar intuition to defeat the trapdoor defense in a white-box setting. 

}

\secspace
\section{Adaptive Attacks}
\label{sec:counter}

In this section, we explore potential adaptive attacks that seek to reduce
the efficacy of \system. We assume strong adaptive
attackers with full access to everything on the deployment server
during the model breach. Specifically, adaptive attackers have:

\begin{packed_itemize} 
\item white-box access to the entire recovery system, 
including the recovery methodology and the GAN used;
\item access to a dataset $D_A$, containing $10\%$ of original training data.
\end{packed_itemize}

\noindent We note that the model owner securely stores the training data and any hidden distributions used in recovery elsewhere offline. 

The most effective adaptive attacks would seek to reduce attack overfitting,
\ie reduce the optimality of the generated attacks w.r.t to the breached
models, since this is the key intuition of \system{}. \emily{However,} these
adaptive attacks must \emily{still produce} adversarial examples that
transfer.
\emily{Thus attackers must strike a delicate balance: using the breached
  models' loss surfaces to search for an optimal attack that would have a high
  likelihood to transfer to the deployed model, but not ``too optimal,'' lest
  it overfit and be detected.}


We consider two general adaptive attack strategies. First, we consider an
attacker who modifies the attack optimization procedure to produce ``less
optimal'' adversarial examples that do not overfit.  Second, we consider ways
an attacker could try to mimic \system\/ by generating its own local model
versions and optimize adversarial examples on them. We discuss the
two attack strategies in \S\ref{sec:reduce-overfit} and
\S\ref{sec:modify-leak} respectively.

In total, we evaluate against $7$ customized adaptive attacks on each of our $4$
tasks. For each experiment, we follow the recovery system setup discussed in
\S\ref{sec:eval}.  When the adaptive attack involves the adaption of existing
attack, we use PGD attack because it is the attack that \system{}
\emily{performs the worst against.} 

\begin{figure*}[t]
\centering

\begin{minipage}{0.3\textwidth}
  \centering
  \includegraphics[width=1\textwidth]{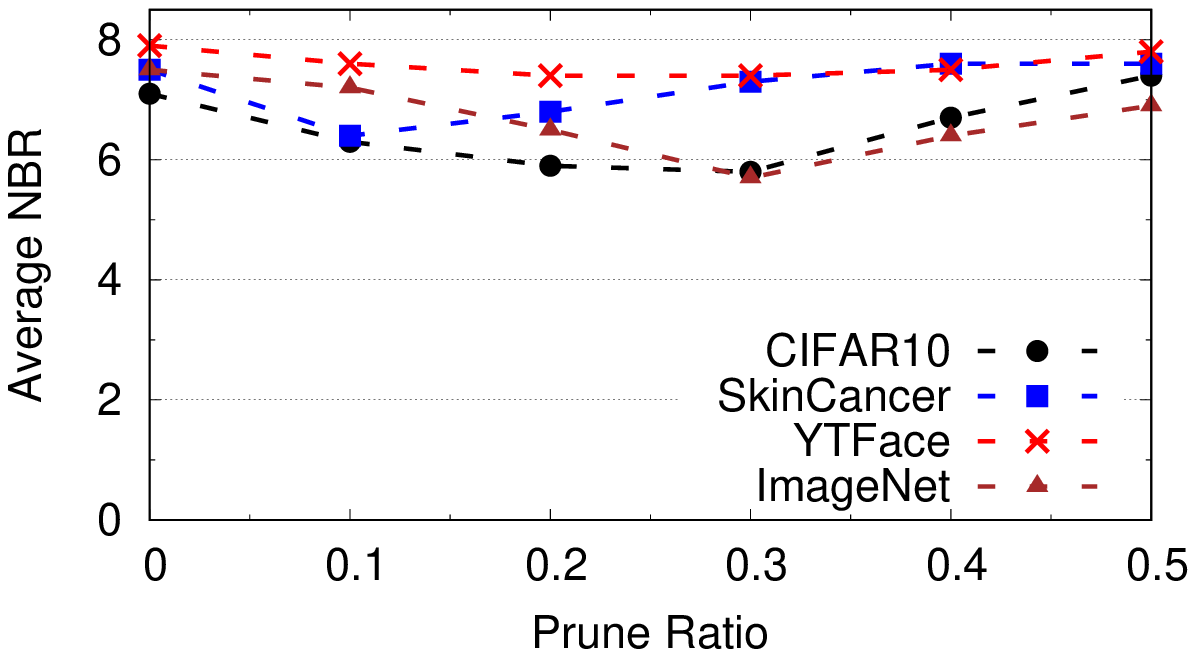}
  \caption{Against adaptive attack that prune $F$ and then finetuning, \system's average NBR decreases then slowly increases as the pruning ratio increases. }
  \label{fig:prune}
\end{minipage}
\hfill
\begin{minipage}{0.3\textwidth}
  \centering
  \includegraphics[width=1\textwidth]{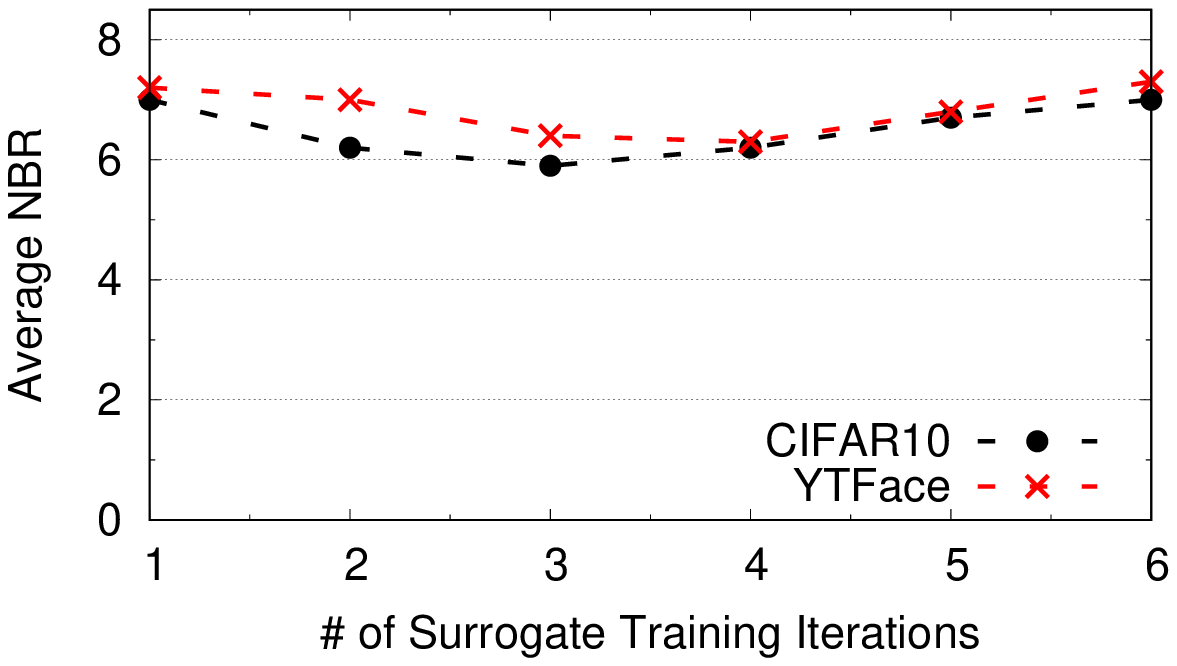}
  \caption{For surrogate model attack, average NBR of \system{}
  decreases then increases as the number of surrogate training 
  iterations increases. }
  \label{fig:surrogate}
\end{minipage}
\hfill
\begin{minipage}{0.34\textwidth}
  \centering
  \resizebox{1.0\textwidth}{!}{
    \begin{tabular}{lcccc}
    \toprule
\multicolumn{1}{c}{\multirow{2}{*}{\textbf{Tasks}}} & \multicolumn{3}{c}{\textbf{Average NBR}} \\ \cline{2-4} 
\multicolumn{1}{c}{} & \textbf{PGD} & \textbf{CW} & \textbf{EAD} \\ \hline
    \cifar & $5.4 \, (\downarrow 1.7) $ & $7.8 \, (\downarrow 1.3) $ & $7.5 \, (\downarrow 1.2) $ \\
    \skin  & $8.3 \, (\uparrow 0.8) $ & --- & --- \\
    \ytface  & $6.4  \, (\downarrow 1.5) $ & $9.9 \, (\downarrow 1.0) $ & $9.1 \, (\downarrow 0.9) $ \\
    \imagenet  & $6.2 \, (\downarrow 1.3) $ & $8.8 \, (\downarrow 0.8) $ & $8.6 \, (\downarrow 1.1) $ \\ \toprule
    \end{tabular}
  }
  \caption{\emily{\system's average NBR remains high against attacks that generate local model versions via unlearning}.}
  \label{tab:gen-new-version}
\end{minipage}
\vspace{-0.25cm}
\end{figure*}

\begin{table}[t]
  \centering
  \resizebox{0.5\textwidth}{!}{
    \begin{tabular}{lcccc}
    \toprule
\textbf{\begin{tabular}[c]{@{}l@{}}Augmentation \\ Method\end{tabular}} & \textbf{\cifar} & \textbf{\skin} & \textbf{\ytface} & \textbf{\imagenet} \\ \midrule
    DI$^2$-FGSM               & $6.6 \, (\downarrow 0.5)$ & $6.7 \, (\downarrow 0.8)$ & $7.3 \, (\downarrow 0.6)$ & $7.0 \, (\downarrow 0.5)$ \\
    VMI-FGSM                  & $6.3 \, (\downarrow 0.8)$ & $6.6 \, (\downarrow 0.9)$ & $7.0 \, (\downarrow 0.9)$ & $6.5 \, (\downarrow 1.0)$ \\
    Dropout ($p = 0.1$)  & $6.5 \, (\downarrow 0.6)$ & $7.0 \, (\downarrow 0.5)$ & $7.2 \, (\downarrow 0.7)$ & $6.9 \, (\downarrow 0.6)$ \\
    Dropout ($p = 0.2$)  & $6.4 \, (\downarrow 0.7)$ & $7.0 \, (\downarrow 0.5)$ & $7.3 \, (\downarrow 0.6)$ & $7.1 \, (\downarrow 0.4)$ \\ \bottomrule
    \end{tabular}
  }
  \caption{\emily{\system's average 
  NBR of remains high against adaptive PGD attacks that leverage different types of data augmentation}. $\downarrow$ and $\uparrow$ denote the decrease/increase in NBR compared to without adaptive attack. }
  \label{tab:augmentation}
  \vspace{-0.5cm}
\end{table}

\begin{table}[t]
  \centering
  \resizebox{0.5\textwidth}{!}{
    \begin{tabular}{lcccc}
    \toprule
    \textbf{\begin{tabular}[c]{@{}l@{}}Target Output \\ Probability\end{tabular}} & \textbf{\cifar} & \textbf{\skin} & \textbf{\ytface} & \textbf{\imagenet} \\ \midrule
    0.9 & $6.9 \, (\downarrow 0.2)$ & --- & --- & --- \\
    0.95 & $6.7 \, (\downarrow 0.4)$ & --- & $7.1 \, (\downarrow 0.8)$ & $6.9 \, (\downarrow 0.6)$\\
    0.99 & $7.0 \, (\downarrow 0.1)$ & $7.3 \, (\downarrow 0.2)$ & $7.6 \, (\downarrow 0.3)$ & $7.7 \, (\uparrow 0.2)$ \\ \bottomrule
    \end{tabular}
  }
  \caption{\system's average NBR  remains high against 
  low-confidence attacks with varying target output probability. ``---'' denotes the 
  attack has $< 20\%$ transfer success rate. } 
\vspace{-0.5cm}
  \label{tab:low-confidence}
\end{table}

\subsection{Reducing Overfitting}
\label{sec:reduce-overfit}

The adaptive strategy here is to intentionally find less optimal
(e.g. weaker) adversarial examples to reduce overfitting. However,
these \emily{less optimal attacks can have low transferability.}  We evaluate
$4$ adaptive attacks \emily{that employ} this strategy. Overall, \emily{we
  find that these types of} adaptive attacks have limited efficacy, reducing
the performance of \system{} by at most $1$ NBR.


\para{Augmentation during attack optimization.} Data augmentation is an effective
technique to reduce overfitting. Recent work~\cite{gao2022boosting,xie2019improving,
byun2022improving,wang2021enhancing} leverages data augmentation to improve the transferability
of adversarial examples. We evaluate \system{} against \rev{five} data 
augmentation approaches, \emily{which are applied} at each attack optimization 
step: 1) DI$^2$-FGSM attack~\cite{xie2019improving} 
which uses series of image augmentation \eg image 
resizing and padding, 2) VMI-FGSM attack~\cite{wang2021enhancing}, 
which leverages more sophisticated image augmentation,   
3) a dropout augmentation approach~\cite{srivastava2014dropout}
where a random portion ($p$) of pixels are set to zero. 

Augmented attacks slightly degrade \system's recovery performance, but the
$NBR$ reduction is limited ($< 0.9$, see Table~\ref{tab:augmentation}).  Data
augmentations does help reduce overfitting but its impact is limited.



\para{Weaker adversarial attacks. } As shown in \S\ref{sec:multi-same},
\system{} achieves better performance on stronger attacks because stronger
attacks overfit more on the breached models, making them easier to
detect. Thus, attackers can test if \textit{weaker} attacks can degrade
\system's performance. We test against two weak adversarial attacks,
SPSA~\cite{uesato2018adversarial} and
DeepFool~\cite{moosavi2016deepfool}. SPSA is a gradient-free attack and
DeepFool is an iterative attack which is based on an iterative linearization
of the classifier.  Both attacks often have much lower attack success than
attacks such as PGD and CW attacks~\cite{shan2020gotta}.

\emily{These weaker attacks degrade our filter performance, but do not
  significantly reduce \system's NBR due to their low
  transferability}. Overall, \system{} maintains $\geq 6.2$ NBR against SPSA
and Deepfool attacks across $4$ tasks. In our tests, both SPSA and Deepfool
attacks have very low transfer success rates ($< 12\%$) on \skin, \ytface, and
\imagenet, even when jointly optimized on multiple breached
versions. \emily{Attacks transfer better on \cifar{} ($37\%$ on
  average), as observed previously, but \system{} still detects nearly
  $70\%$ of successfully transferred adversarial examples. } 


\para{Low confidence adversarial attack. } Another \emily{weak attack is a}
``low confidence'' attack, where the adaptive attacker ensures attack
optimization does not settle in any local optima.  \emily{To do this}, the
attacker constructs adversarial examples that do not have $100\%$ output
probability on the breached versions (over $97\%$ of all PGD adversarial
examples reach $100\%$ output probabilities).

Table~\ref{tab:low-confidence} shows the NBR of \system{} against
low-confidence attacks with an increasing target output
probability. \emily{Low confidence attacks tend to produce attack samples
  that do not transfer, \eg ineffective attack samples.  For samples that
  transfer better, \system{} maintains a high NBR ($\geq 6.7$) across all
  tasks. }


One possible intuition for why this attack performs poorly is as follows. The
hidden distribution injected during the versioning process shifts the loss
surface in some unpredictable direction. Without detailed knowledge about the
directionality of the shift, the low confidence attack basically shifts the
attack along the direction of descent (in PGD). If this directional vector
matches the directionality of the shift introduced by \system{}, then it
could potentially reduce the loss difference $\Delta_{max}$. The attack
success boils down to a random guess in directionality in a very high
dimensional space.


\para{Moving adversarial examples to sub-optimal locations.}
\emily{Next, we try an advanced approach in which we move adversarial
  examples away from the local optima, and search for an adversarial example
  whose loss is different from the local optima exactly equivalent to the
  loss difference value used by our filter for detection. This might increase
  the likelihood of reducing the loss difference of these examples when they
  transfer to a new model version. We assume the attacker can use iterative
  queries to probe and determine the threshold value $T_{i+1}$ (\S\ref{sec:method}).}



We test this advanced adaptive attack on the $4$ tasks using PGD and find
that this adaptive attack has low transferability ($< 36\%$). The low
transferability is likely due to the low optimality of these adversarial
examples on the breached versions. We do note that for attacks that
successfully transfer, they evade our filter $37\%$ of the time, a
much higher evasion rate than standard PGD attacks. Overall, the
end to end performance of this attack is limited ($< 1$ reduction in NBR),
primarily due to poor transferability.

\rev{

\para{Logit matching attack. } A logit matching attack~\cite{sabour2015adversarial} matches the feature space
representation of the adversarial examples with target feature presentations. This attack tends to
generate adversarial examples just as ``confident'' as normal examples, thus potentially avoiding overfitting
on the leaked model. We test the logit matching attack on all $4$ datasets and found that the attacks have very low
transferability (< $32\%$). For those attacks that do transfer successfully, \system{} detects $92\%$ of them. 
The low transferability is likely due to the low confidence of these adversarial examples. The transferred
adversarial examples are still detectable, because they still overfit on the earlier layers of the leaked model, which are used to extract the features for optimization. 

}

\vspace{-0.15in}
\subsection{Modifying breached Versions}
\label{sec:modify-leak}

\looseness=-1 Here, the attackers try a different strategy, and try to generate their own
local ``version'' of the model. \emily{The attacker hopes to construct
  adversarial examples that may overfit on the local version but not the
  breached version, thus evading detection.}  This type of adaptive attack
faces a similar tradeoff as before.  To generate a local version $F'$,
attacker must leverage information from the breached model versions because
they do not have enough training data to train from scratch.  Yet,
\emily{leveraging breached versions} means that $F'$ may have a similar loss
surface to the breached versions, causing adversarial examples to still
overfit on the breached version and be detected.

We evaluate $3$ adaptive attacks that use different mechanisms to generate a
new $F'$ from the original breached versions. In case of multiple breached
versions, attacker applies adaptive attacks on each version to generate 
$F_1',...,F_i'$ and jointly optimizes adversarial examples. Overall, these
attacks have limited efficacy, reducing average NBR by $\le 1.7$.

\para{Finetuning with benign data.} A simple approach to generate $F'$ is to
directly finetune each breached version on the \emily{attacker's} small set
of training data ($D_A$). However, directly finetuning on benign data has
limited impact on the original breached versions and thus, limited impact on
\system{} (see the Appendix).  To increase the impact
of finetuning, we ``prune'' the weights of breached versions before
retraining \emily{by randomly setting some weights to zero}.  We then retrain
the pruned model on $D_A$ to produce $F'$. The attacker can control the
impact \emily{of pruning on} $F$ by \emily{changing the ``pruning ratio''
  (proportion of weights pruned).}

We test this adaptive attack on all $4$ tasks using 
PGD attacks on $F'$. Figure~\ref{fig:prune}
shows the NBR of \system{} decreases gradually to $5.5$ as pruning ratio
increases to $0.3$, showing the adaptive attack is effective. However, when
\emily{pruning ratio $\ge 0.3$, the average NBR} of \system{} returns to its
original level. \emily{This is because attack transferability decreases as
  $F'$ becomes increasingly different (due to higher pruning ratio) from the
  breached/new versions.}   

\para{Surrogate model attack. } Next, we consider an adaptive attack who
trains a local version from scratch using techniques borrowed from ``model
stealing'' attacks~\cite{papernotblackbox}. As stated in \S\ref{sec:define},
we do not consider surrogate model stealing attack against the new version
due to effective server-side defenses.  In our test, we implement the
surrogate model training technique from~\cite{papernotblackbox}, which
iteratively trains a surrogate model by querying the breached versions.
\emily{The model stealing attack only produces high performing model
  surrogate models for \cifar{} and \ytface, so we restrict our evaluation to
  these tasks. Surrogate attacks are unsuccessful on \skin{} and \imagenet{}
  datasets, \ie $< 2\%$ transfer success rate.  This is unsurprising, since \skin{}
  and \imagenet{} are challenging to learn even with the full
  dataset.}

Against PGD attacks generated on these surrogate versions, \system{} has a
high filter success rate ($> 94.9\%$ when attacker breaches $1$ version) . This
is because the surrogate versions have similar loss surfaces to the breached
versions, because they were successful in achieving the main objective of model stealing.
Figure~\ref{fig:surrogate} shows the NBR of \system{} as attacker trains the
surrogate with an increasing number of iterations. The average NBR of
\system{} decreases (by $\le 1.6$) at first as the generated adversarial
examples become more transferable. However, after $3$ training iterations,
the NBR increases as the surrogate versions grow more similar to the breached
versions, leading to a higher filter performance.

More recent work on model stealing attacks~\cite{yuan2022attack,yu2020cloudleak}
claim even stronger ability to duplicate the target model's classification
surface (compared to \cite{papernotblackbox}). However, this makes these
attacks even more similar to the breached model versions, and therefore even
easier to detect by \system's filter.

\para{Generating local version via unlearning and retraining. } This adaptive attack explores the 
possibility of attacker generating a local version $F'$ that is indistinguishable from any possible version generated by \system. If this is possible, adversarial examples optimized on such $F'$ should transfer to any breached and new versions with a small $\Delta_{max}$. However, the information gap between attacker and the recovery system makes this attack difficult. 
Using only the breached version and limited training data, the attack must 1)
remove the original hidden distributions injected by \system, and 2) inject
new hidden distributions.  Existing work on machine
unlearning~\cite{bourtoule2021machine,guo2019certified} shows that completely
``unlearning'' a subset of training data is very challenging. To make the
problem even harder, the attacker does not know but must correctly guess the
exact hidden distributions injected by \system.

\emily{Thus, we assume} attacker uses an unlearning method~\cite{vyas2018out,shan2022poison} to
unlearn the entire GAN output data distribution from the breached version, hoping that in the process it unlearns the original hidden distributions. 
After the unlearning process converges, attacker trains in new hidden distributions using \system's methodology. 


On \cifar, \ytface, and
\imagenet, this adaptive attack slightly decreases \system's performance ($< 1.7$ decrease in average NBR, see Table~\ref{tab:gen-new-version}).  
The limited impact is likely due to the inability to fully unlearn the effect of original hidden distributions. 
On \skin, this adaptive attack performs \textit{worse} than the standard attacks. This is because unlearning 
significantly modifies the loss surface of the original
model, leading to adversarial examples with 
poor transferability. The smaller size ($50K$ images) and the more challenging learning task (low benign accuracy) of \skin{} dataset also make unlearning more challenging for the adaptive attacker.

\rev{
\section{Limitations}

\para{Threat of adaptive attacks.} Despite our best efforts to design and evaluate 
potential adaptive attacks, it is likely that more advanced adaptive 
attacks could be designed to
bypass our system. We leave the design and evaluation of stronger 
adaptive attacks against \system{} as future work. 

}

\para{Deployment of all previous versions in each filter.} To calculate the
detection metric $\Delta_{max}(x)$, filter $D_{i+1}$ includes all previously
breached models ($F_1 \ldots F_i$) alongside $F_{i+1}$. This
has two implications. First, if an attacker later breaches version $i+1$,
they automatically gain access to all previous versions.  This simplifies the
attacker's job, making it faster (and cheaper) for them to collect
multiple models to perform ensemble attacks. Second, the filter induces an
inference overhead as inputs now need to go through each previous
version. While this can be parallelized to reduce latency, total inference
computation overhead grows linearly with the number of breaches.

We also considered an alternative design for \system, where we do not use
previously breached models at inference time. Instead, for each input, we use local gradient search to find any nearby local loss minima, and use it to approximate the amount of potential overfit to a previously breached
model version (or surrogate model) ($\Delta_{max}(x)$ in eq.(\ref{eq:lossdiff})). While it avoids the limitations listed 
above, this approach relies on simplifying assumptions of the minimum loss
value across model versions, which may not always hold. In addition, it
requires multiple gradient computations for each model input, making it
prohibitively expensive in practical settings.

\looseness=-1 \para{Limited number of total recoveries possible.} \system{}'s ability to
recover is not unlimited. It degrades over time against an attacker with an
increasing number of breached versions. This means \system{} is no longer
effective once the number of actual server breaches exceeds its NBR. While
current results show we can recover after several server breaches even under
strong adaptive attacks (\S\ref{sec:counter}), we consider this work as an 
initial step, and expect future solutions that can provide even
stronger recovery properties.

\vspace{-0.1in}
\section{Conclusion}
\label{sec:discussion}

This work identifies the model recovery problem and proposes an initial
solution, \system. \system{} introduces small, unpredictable shifts in the
classification surface between different model versions it produces, making
it possible to identify adversarial examples generated on leaked models
because of their tendency to overfit.
\system{} achieves high performance (restores model functionality following a significant number of server breaches) under a
variety of scenarios. The strongest adaptive attacks we can design only decrease its NBR by a small amount.

Our work is an initial step towards addressing the difficult challenge of
recovery after a model leak. We hope our work motivates follow-on 
systems that provide significantly stronger properties than our own.

\section*{Acknowledgements}  

We thank our anonymous reviewers and shepherd for their insightful 
feedback. This work is supported in part by NSF grants CNS1949650, 
CNS-1923778, CNS-1705042, by C3.ai DTI, and by the DARPA GARD program. 
Emily Wenger is supported by a GFSD Fellowship, a Harvey Fellowship, and a 
Neubauer Fellowship. Shawn Shan is supported by an Eckhardt
Fellowship at the University of Chicago. Any opinions, findings, and 
conclusions or recommendations expressed in this material are those of the 
authors and do not necessarily reflect the views of any funding agencies.


\bibliographystyle{ACM-Reference-Format}
\bibliography{salt}
\balance

\appendix
\section{Appendix}
\label{sec:appendix}

\begin{table}[t]
  \centering
  \resizebox{0.5\textwidth}{!}{
  \begin{tabular}{lcccc}
  \toprule
  \textbf{Tasks} & \textbf{Input Size} & \textbf{\# Classes} & \textbf{\# Training Data} & \textbf{Architecture} \\ \midrule
  \cifarS & $32 \times 32$ & $10$ & $50,000$ & ResNet-18~\cite{he2016deep} \\
  \skin & $224 \times 224$ & $7$ & $50,000$ & ResNet-101~\cite{he2016deep} \\
  \ytface & $224 \times 224$ & $1,283$ & $375,645$ & ResNet-101~\cite{he2016deep} \\
  \imagenet & $299 \times 299$ & $1,000$ & $1,281,167$ & Inception ResNet~\cite{szegedy2017inception} \\ \bottomrule
  \end{tabular}
  }
  \caption{Datasets \& DNN architectures for our evaluation.}
  \label{tab:tasks}
  \vspace{-0.5cm}
\end{table}

\begin{table}[t]
  \centering
  \resizebox{0.49\textwidth}{!}{
\begin{tabular}{lcccc}
\hline
\textbf{Model} & \textbf{Optimizer} & \textbf{\# Epochs} & \textbf{Batch Size} & \textbf{Start learning rate} \\ \hline
CIFAR & SGD & 100 & 512 & 0.1 \\
Skin & Adam & 50 & 32 & 0.005 \\
YTF & Adam & 50 & 32 & 0.005 \\
ImageNet & Adam & 100 & 32 & 0.005 \\ \hline
\end{tabular}
  }
  \caption{Detailed information on our model training configurations.}
  \label{tab:training-params}
\end{table}

\begin{figure}
  \centering
  \includegraphics[width=0.85\linewidth]{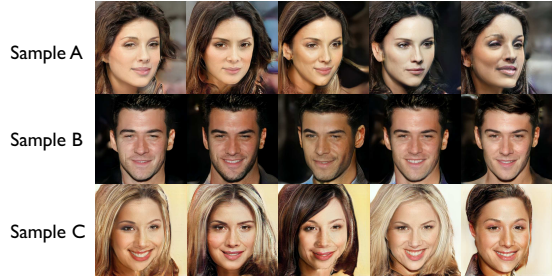}
  \caption{Example images generated from PGAN. Each row corresponding to a
  different hidden distribution. } 
  \label{fig:gan_samples}
\end{figure}

\begin{figure*}[t]
\centering

  \begin{minipage}{0.32\textwidth}
  \centering
  \includegraphics[width=1\textwidth]{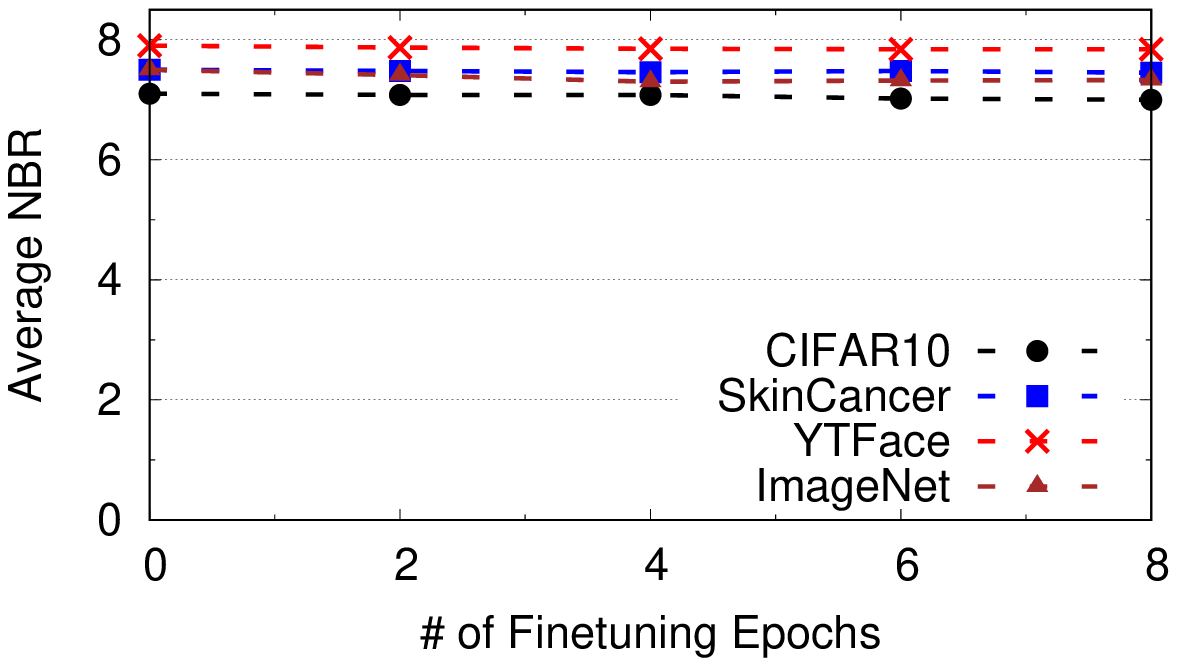}
  \caption{For adaptive attack that finetunes $F$ on benign data, average NBR of our system 
  slightly decreases as the number of finetuning epochs increases. }
  \label{fig:benign-tune}
\end{minipage}
\hfill
\begin{minipage}{0.32\textwidth}
\centering
\includegraphics[width=1\textwidth]{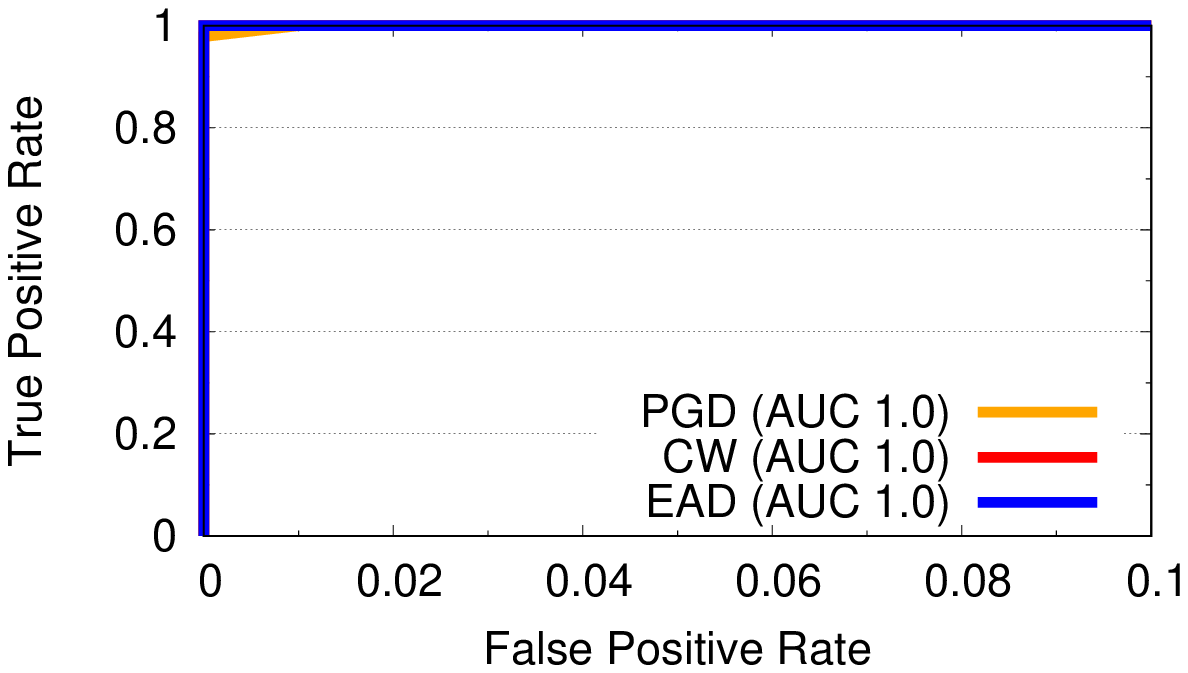}
\caption{ROC curve of detecting adversarial examples generated using the $3$ adversarial attacks 
on \cifar. The x-axis is concatenated at 0.1 for brevity. (Single Leakage)}
\label{fig:auc-cifar}
\end{minipage}
\hfill
  \begin{minipage}{0.32\textwidth}
  \centering
  \includegraphics[width=1\textwidth]{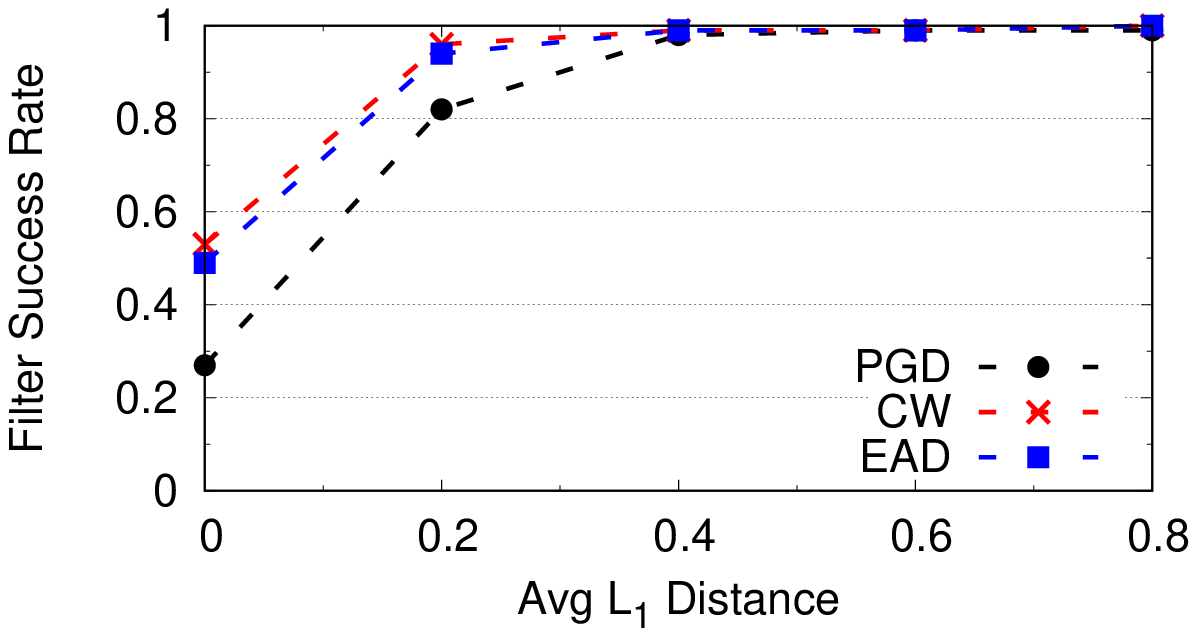}
  \caption{Analysis on the impact of selecting 
  similar hidden distributions. The filter success rate increase rapidly 
  as the $L_1$ distance between sampling Gaussian distributions increases. }
  \label{fig:collision}
  \end{minipage}
\end{figure*}

\subsection{Detailed Proof of Theorem~\ref{thm:loss}}
\label{appendix:proof}
We now present the detailed proof of Theorem~\ref{thm:loss} described
in \S\ref{sec:theory}.  We start from presenting detailed definitions of
attack optimization, transferability, and assumptions of $F$ and $G$,
and then discuss how $G$ and its detector respond to an adversarial
example computed from $F$. 

\para{Attack Optimization on $F$.} Let $x'=x + \delta$ be an adversarial example optimized from model $F$ with target label $y_t$. We approximate $\ell_2$ loss around $x'$ as loss of a linear classifier.
\begin{equation}
  \ell_2(F(x'), y_t) = (y_t-(a_F x' + b_F))^2
\end{equation}
We consider the perturbation $\delta$ to be optimized on $F$ for
benign input $x$. Therefore, $\ell_2(F(x'), y_t) \le \gamma$ for some
small $\gamma > 0$. This yields $\frac{y_t-b_F-\sqrt{\gamma}}{a_F} \le
x' \le \frac{y_t-b_F+\sqrt{\gamma}}{a_F}$.  

\para{Attack Transferability to $G$.}  Let $G$ be a model being attacked by the adversarial example $(x', y_t)$. Similarly, the loss around $x'$ on $G$ is approximated by
\begin{equation}
  \ell_2(G(x'), y_t) = (y_t-(a_G x' + b_G))^2.
\end{equation}
The adversarial example $x'$ transfers to $G$ if its loss with respect
to $y_t$ is bounded by some $\gamma'$ such that
$\ell_2(G(x'), y_t) \le \gamma'$. This yields
$\frac{y_t-b_G-\sqrt{\gamma'}}{a_G} \le x' \le
\frac{y_t-b_G+\sqrt{\gamma'}}{a_G}$. Note that since $x'$ is not
optimized on $G$, naturally $\gamma' >> \gamma$.

\para{Assumptions on $F$ and $G$.}  Since $F$ and $G$ are trained
using the same benign training data, we assume $a_F = a_G=a$. On the
other hand, $b_F \neq b_G$ due to the different hidden
distributions. Without loss of generality, we assume $a>0$ and $b_G >
b_F$.  Our result remains the same when $b_G<b_F$.

\para{How model $G$ and its attack detector respond to $x'$.}  There
are 
two cases. 


\noindent {\bf Case 1:} 
If $b_G - b_F > \sqrt{\gamma'} - \sqrt{\gamma}$, then 
$[\frac{y_t-b_G-\sqrt{\gamma'}}{a_G},
\frac{y_t-b_G+\sqrt{\gamma'}}{a_G}]$ does not fully contain 
$[\frac{y_t-b_F-\sqrt{\gamma}}{a_F},
\frac{y_t-b_F+\sqrt{\gamma}}{a_F}]$ and the adversarial example $x'$
will not tranfer to $G$.  In this case, $x'$ can be easily
identified since $F(x') \neq G(x')$.

\noindent {\bf Case 2:}  If $ b_G - b_F \le \sqrt{\gamma'} -
\sqrt{\gamma}$, then $[\frac{y_t-b_F-\sqrt{\gamma}}{a_F},
\frac{y_t-b_F+\sqrt{\gamma}}{a_F}]$ is fully contained in
$[\frac{y_t-b_G-\sqrt{\gamma'}}{a_G},
\frac{y_t-b_G+\sqrt{\gamma'}}{a_G}]$. In this case, $x'$ can cause 
$G(x')=y_t$. 
This is when we use the filter defined by eq.(\ref{eq:lossdiff}) to compare the loss of $(x',y_t)$ between $F$ and $G$.

We now study $\ell_2(G(x'), y_t) -
\ell_2(F(x'), y_t)$. Since $a_F = a_G=a$, we can expand the
expression as: 

\begin{align*}
    & \ell_2(G(x'), y_t) - \ell_2(F(x'), y_t) \\
    = & (y_t - (a x' + b_G))^2 - (y -(a x' + b_F))^2 \\
    = & 2y_t(a x' + b_F) - 2y_t(a x' + b_G) + (a x' + b_G)^2 - (a x' + b_F)^2 \\
    = & 2y_t(b_F - b_G) + 2a x'(b_G - b_F) + b_G^2 - b_F^2 
\end{align*}
Since $x'$ lies in the interval of
$[\frac{y_t-b_F-\sqrt{\gamma}}{a},
\frac{y_t-b_F+\sqrt{\gamma}}{a}]$,  we assume $x'$ is uniformly
distributed in this interval.  Thus we have 
\begin{align*}
  & \Pr(\ell_2(G(x'), y_t) - \ell_2(F(x'), y_t) > T) \\
  = & \Pr(2y_t(b_F - b_G) + 2a x'(b_G - b_F) + b_G^2 - b_F^2 > T) \\
  = & \Pr(x' > \frac{T + b_F^2 - b_G^2 - 2y_t(b_F - b_G)}{2a (b_G - b_F)} ) \\
  = & \frac{a}{2\sqrt{\gamma}} \left( \frac{y_t-b_F+\sqrt{\gamma}}{a} - \frac{T + b_F^2 - b_G^2 - 2y_t(b_F - b_G)}{2a (b_G - b_F)} \right) \\
  = & \frac{a}{2\sqrt{\gamma}} \cdot \frac{(b_G - b_F)(b_G - b_F + 2\sqrt{\gamma}) - T}{2a (b_G - b_F)} \\
  = & \frac{(b_G - b_F)(b_G - b_F + 2\sqrt{\gamma}) - T}{4 \sqrt{\gamma} (b_G - b_F)}
\end{align*}
 Let $\Pr(\ell_2(G(x'), y_t) - \ell_2(F(x'), y_t) > T) =p$,  which
 yields $T = (b_G-b_F)(b_G-b_F+2\sqrt{\gamma}-4\sqrt{\gamma}p)$.  Now let $\mathcal{D}_{G,F}= b_G-b_F$, we have $T = \mathcal{D}_{G,F}
\cdot (\mathcal{D}_{G,F} + 2\sqrt{\gamma} - 4\sqrt{\gamma} \cdot p)$. 

\noindent This completes the proof of Theorem~\ref{thm:loss}. 

\para{Discussion.} Note that {\bf Case 2} must meet the constraint of  $b_G - b_F \le \sqrt{\gamma'} -
\sqrt{\gamma}$.  Thus if we train $F$ and $G$ to be ``well-separated''
when classifying non-benign inputs,  we can set $b_G - b_F = \sqrt{\gamma'} -
\sqrt{\gamma}$. If $p=0.95$, $\frac{\gamma'}{\gamma}=25$, then
$T=8.8\gamma$.

\begin{table}[t]
  \centering
  \resizebox{0.5\textwidth}{!}{
  \begin{tabular}{lcccc}
  \toprule
  \textbf{Iterations} & \textbf{\cifar} & \textbf{\skin} & \textbf{\ytface} & \textbf{\imagenet} \\ \midrule
  1 & 7.3 & --- & ---  & 7.7 \\
  10 & 6.9 & 7.2 & 7.7 & 7.4 \\
  50 &  7.1 & 7.5 & 7.8 & 7.5 \\
  100 & 7.1 & 7.5 & 7.9 & 7.5 \\ \bottomrule
  \end{tabular}
  }
  \caption{Average NBR of \system{} against PGD attack first decrease 
  and then increases slightly as the optimization iterations 
  of PGD attack increases. }
  \label{tab:iteration-impact}
\end{table}

\subsection{Implementation Details}
\label{sec:implement-details}
\label{sec:hidden-gen}

\para{Soft Nearest Neighbor Loss Term. } For each label in $L$, we calculate the SNNL loss term 
as the following: 

\begin{equation}
SNNL(X,Y)=-\frac{1}{N}\sum_{i\in 1..N}^{ }log\left(\frac{\sum\limits_{\substack{j\in 1..N \\ j\neq i \\ y_{i} = y_{j}}}^{ }e^{-||x_{i} - x_{j}||^{2}}}{\sum\limits_{\substack{k\in 1..N \\ k\neq i}}^{ }e^{-||x_{i} - x_{k}||^{2}}}\right) 
\label{eq:snnl}
\end{equation}

\noindent $N$ is the total number of training data, $x_i$ is the i-th training data. The equation effectively minimize the distance of training data that are from the same dataset, while maxmizing distance of training data that are different classes.

\para{Generating Hidden Distribution using GAN}. We use a 
GAN~\cite{karras2017progressive} trained on CelebA dataset~\cite{liu2018large}. 
The GAN takes in an input vector of size $512$ 
and output a $224 \times 224$ facial image.\footnote{We use an image generation 
GAN in this paper as we consider only computer vision tasks. However, GANs for other 
domains are also available, and we leave the application of our recovery system to 
other domains as future work.} The GAN is trained with input vectors sampled from
a Gaussian ball $\mathcal{N}(\mu=0,\,\sigma^{2}=1)$. For each hidden distribution, we sample from a 
smaller Gaussian ball within the original Gaussian ball that has a random mean vector 
(bounded between $-0.5$ and $0.5$) and a small standard deviation ($\sigma_0$). Then we query the 
GAN using noise vector sampled from the smaller Gaussian ball and take the generated images as the current 
hidden distribution. We empirically 
choose $\sigma_0 = 0.3$, which generates similar images but has enough variety  
as shown in Figure~\ref{fig:gan_samples} in Appendix. In \S\ref{sec:eval-single}, we show that there
exist a large number of hidden distributions in the GAN for our version generation purpose.

Next, we seek to estimate the total number 
of different hidden distributions exist in the GAN. To do so, we find the minimal 
separation that two versions needs to have 
in order to achieve high filter success rate. We train
versions $F_0$ and $F_1$ using extremely similar hidden distributions
(measured by the distance between the mean of Gaussian distributions used
to sample the hidden distributions). 

Figure~\ref{fig:collision} shows the
filter success rate as the $L_1$ distance between the mean of 
the input Gaussian distributions (\eg 512-long vectors) used to sample $F_0$ 
and $F_1$'s hidden distributions increases. We see that as long as $L_1$ 
distance between the sampling Gaussian distribution is 
above $0.4$, \system{} can maintain $>98\%$ filter success rate. 
For reference, the maximum $L_1$ distance between the mean of two input Gaussian distribution
is $512$. Note that even when $F_0$ and $F_1$ uses the exact same hidden 
distributions ($L_1 = 0$) the filter success rate is higher than zero. This is because the stochasticity 
of floating point GPU computation causes versions to be slightly different even 
the training data and parameter initialization is identical. 
Since our space of input Gaussian distribution is
large ($512$ dimensions, each continuous from $-0.5$ to $0.5$), 
we have a large number of possible hidden distributions ($\geq 2^{512}$).

\end{document}